\documentclass[trackchanges,twocolumn]{aastex7}
\usepackage{multirow}

\newcommand{\code}[1]{\texttt{#1}}

\newcommand\msun{\text{M}_{\odot}}
\newcommand\teff{\text{T}_{\text{eff}}}
\newcommand\mjup{\text{M}_{\text{Jup}}}

\begin{document}

\title{Probing the Outskirts of M Dwarf Planetary Systems with a Cycle 1 JWST NIRCam Coronagraphy Survey}

\author[0000-0002-7325-5990]{Ellis Bogat}
\affiliation{Department of Astronomy,
University of Maryland,
College Park, MD 20782, USA}
\affiliation{NASA Goddard Space Flight Center,
8800 Greenbelt Rd, 
Greenbelt, MD 20771, USA}
\affiliation{Center for Research and Exploration in Space Science and Technology, NASA/GSFC, Greenbelt, MD 20771}
\email{ebogat@umd.edu}

\author[0000-0001-5347-7062]{Joshua E. Schlieder}
\affiliation{NASA Goddard Space Flight Center,
8800 Greenbelt Rd, 
Greenbelt, MD 20771, USA}
\email{joshua.e.schlieder@nasa.gov}

\author[0000-0002-6964-8732]{Kellen D. Lawson}
\affiliation{NASA Goddard Space Flight Center,
8800 Greenbelt Rd, 
Greenbelt, MD 20771, USA}
\email{kellen.d.lawson@nasa.gov}

\author[0000-0002-6845-9702]{Yiting Li}
\affiliation{Department of Astronomy, University of Michigan, 1085 S. University, Ann Arbor, MI 48109, USA}
\email{lyiting@umich.edu}

\author[0000-0002-0834-6140]{Jarron M. Leisenring}
\affiliation{Steward Observatory, 
University of Arizona, 
933 N. Cherry Avenue, 
Tucson, AZ 85721, USA}
\email{jarronl@arizona.edu}

\author[0000-0003-1227-3084]{Michael R. Meyer}
\affiliation{Department of Astronomy, University of Michigan, 1085 S. University, Ann Arbor, MI 48109, USA}
\email{mrmeyer@umich.edu}

\author[0000-0001-6396-8439]{William Balmer}
\affiliation{Johns Hopkins University,
Baltimore, MD, USA}
\email{wbalmer@stsci.edu}

\author[0000-0001-7139-2724]{Thomas Barclay}
\affiliation{NASA Goddard Space Flight Center,
8800 Greenbelt Rd, 
Greenbelt, MD 20771, USA}
\email{thomas.barclay@nasa.gov}

\author[0000-0002-5627-5471]{Charles A. Beichman}
\affiliation{NASA Exoplanet Science Institute/IPAC, 
California Institute of Technology, 
1200 E. California Blvd., Pasadena, CA 91125, USA}
\affiliation{Jet Propulsion Laboratory, California Institute of Technology, Pasadena, CA 91011, USA}
\email{charles.a.beichman@jpl.nasa.gov}

\author[0000-0001-5966-837X]{Geoffrey Bryden}
\affiliation{Jet Propulsion Laboratory, California Institute of Technology, Pasadena, CA 91011, USA}
\email{geoffrey.bryden@jpl.nasa.gov}

\author[0000-0002-5335-0616]{Per Calissendorff}
\affiliation{Department of Astronomy, University of Michigan, 1085 S. University, Ann Arbor, MI 48109, USA}
\email{percal@umich.edu}

\author[0000-0001-5365-4815]{Aarynn Carter}
\affiliation{Space Telescope Science Institute, 3700 San Martin Drive, Baltimore, MD 21218, USA}
\email{aacarter@stsci.edu}

\author[0000-0003-1863-4960]{Matthew De Furio}
\affiliation{Department of Astronomy, University of Michigan, 1085 S. University, Ann Arbor, MI 48109, USA}
\email{defurio@umich.edu}

\author[0000-0001-8627-0404]{Julien H. Girard}
\affiliation{Space Telescope Science Institute, 3700 San Martin Drive, Baltimore, MD 21218, USA}
\email{jgirard@stsci.edu}

\author[0000-0002-8963-8056]{Thomas P. Greene}
\affiliation{NASA Ames Research Center, MS 245-6, Moffett Field, CA 94035, USA}
\email{Tom.Greene@nasa.gov}

\author[0000-0001-5978-3247]{Tyler D. Groff}
\affiliation{NASA Goddard Space Flight Center,
8800 Greenbelt Rd, 
Greenbelt, MD 20771, USA}
\email{tyler.d.groff@nasa.gov}

\author[0000-0003-2769-0438]{Jens Kammerer}
\affiliation{European Southern Observatory, Karl-Schwarzschild-Straße 2, 85748 Garching, Germany}
\email{Jens.Kammerer@eso.org}

\author[0000-0002-3414-784X]{Jorge Llop-Sayson}
\affiliation{Jet Propulsion Laboratory, California Institute of Technology, Pasadena, CA 91011, USA}
\email{jllopsay@caltech.edu}

\author[0000-0003-0241-8956]{Michael W. McElwain}
\affiliation{NASA Goddard Space Flight Center,
8800 Greenbelt Rd, 
Greenbelt, MD 20771, USA}
\email{michael.w.mcelwain@nasa.gov}

\author[0000-0002-7893-6170]{Marcia J. Rieke}
\affiliation{Steward Observatory, 
University of Arizona, 
933 N. Cherry Avenue, 
Tucson, AZ 85721, USA}
\email{mrieke@gmail.com}

\author[0000-0001-7591-2731]{Marie Ygouf}
\affiliation{Jet Propulsion Laboratory, California Institute of Technology, Pasadena, CA 91011, USA}
\email{marie.ygouf@jpl.nasa.gov}

\begin{abstract}

The population of giant planets on wide orbits around low-mass M dwarf stars is poorly understood, but the unprecedented sensitivity of JWST NIRCam coronagraphic imaging now provides direct access to planets significantly less massive than Jupiter beyond 10 AU around the closest, youngest M dwarfs. 
We present the design, observations, and results of JWST GTO Program 1184, a Cycle 1 NIRCam coronagraphic imaging survey of 9 very nearby and young low-mass stars at $3 - 5 \mu$m wavelengths. 
In the F356W and F444W filters, we achieve survey median $5\sigma$ contrasts deeper than $10^{-5}$ at a separation of $1^{\prime\prime}$, corresponding to $0.20~\text{M}_{\text{Jup}}$ in F444W and $1.30~\text{M}_{\text{Jup}}$ in F356W at planet-star separations of 10 AU. Our results include $3-5 \mu$m debris disk detections and the identification of many extended and point-like sources in the final post-processed images. 
In particular, we have identified two marginal point source candidates having fluxes and color limits consistent with model predictions for young sub-Jupiter mass exoplanets. 
Under the assumption that neither candidate is confirmed, we place the first direct-imaging occurrence constraints on M dwarf wide-orbit (semimajor axes $10-100$ AU), sub-Jupiter mass exoplanets ($0.3-1~\text{M}_{\text{Jup}}$). 
We find frequency limits of $<0.10$ and $<0.16$ objects per star with 1 and 3$\sigma$ confidence, respectively. 
This survey brings to the forefront the unprecedented capabilities of JWST NIRCam coronagraphic imaging when targeting young, low-mass stars and acts as a precursor to broader surveys to place deep statistical constraints on wide-orbit, sub-Jupiter mass planets around M dwarfs.

\end{abstract}

\section{Introduction}


The population distribution of exoplanet masses, radii, and semi-major axes provides critical insight into the phenomena that govern planet formation and evolution, including those of our own planet and its neighbors. However, the overwhelming majority of planet discoveries have been through the techniques of transit photometry and radial velocity, most sensitive to the inner ($\lesssim$1 AU) regions of planetary systems \citep{foreman-mackey_exoplanet_2014,ananyeva_exoplanets_2023, akeson_nasa_2013}. Gravitational microlensing observations in dense star fields toward the Milky Way bulge provide access to exoplanets at intermediate to wide separations \citep[0.5 - 10 AU,][]{Gaudi2012}, but the nature of the measurement does not allow direct detection and deep characterization of individual planets.

Using the technique of direct imaging with coronagraphy, light originating from the planet can be separated from the starlight, and the sensitivity to exoplanets actually improves with increased planet-star separation. This is achieved by suppressing the starlight with a series of physical masks in the instrument's optical path, in combination with specific observational approaches and image post-processing techniques. Such detections provide access to direct information about the planet's thermal and atmospheric properties. As new and more capable observatories with direct imaging instruments come online, we gain access to widely separated exoplanets in new parts of parameter space (temperature, mass, age, etc.), providing context for the inner planets we have already discovered in large numbers, and enabling further insights into planet formation, evolution, and habitability throughout the galaxy. 


In addition, the observation of young systems ($\lesssim$1 Gyr) allows us to link our knowledge of mature planets and disks with their formation and early history \citep{marleau_constraining_2014,bowler_imaging_2016,meshkat_direct_2017}. Theoretical models generally predict that planets form either by core accretion \citep{pollack1996} or by direct gravitational collapse \citep{boss1997}, and then cool over time until they reach thermal equilibrium with the radiation from the host star and with heat generated by gravitational contraction \citep{chabrier_understanding_2010}. A planet's early temperature evolution ($\lesssim$100 Myr) is highly dependent on its formation mechanism \citep{Spiegel2012}, so direct measurements of the temperature (via direct imaging), mass (via combining with RV or stellar astrometry), and age (via host star or planet properties) are critical for constraining current models \citep{rogers_unveiling_2021}.

M dwarfs ($<0.6~\msun,~\teff \lesssim 4000$ K) are the most common outcome of star formation, by number comprising $\sim70\%$ of the Milky Way's stellar content \citep{Bochanski2010,Henry2024}. Statistics from $Kepler$ led to the surprising revelation that M dwarfs host Earth to Neptune-sized planets in great numbers at small physical separations ($\lesssim$0.5 AU; \citet{Dressing2015, Hardegree2019}). Results from RV surveys support these planet demographics at small separations \citep{Bonfils2013, Mignon2025}, but they also reveal longer-term periodic signals or trends in a few stars, consistent with $\sim$1 M$_{\rm Jup}$ planets at 1--5 AU and a steeper planet mass-function than higher-mass stars \citep{Montet2014, Clanton2016a}. 

Gravitational microlensing surveys have revealed a significant population of Neptune to Jupiter mass planets on $\sim$0.5-10 AU orbits \citep{Cassan2012, Mroz2023}, with the majority of the host stars in the M dwarf mass regime\footnote{$173/235$, or 74\%~of microlensing planet hosts have masses $\le0.6~\msun~$(NASA Exoplanet Archive, queried February 22, 2025)}. The estimated occurrence rates of these planets are also consistent with a steep mass-function where Jovian planets are relatively rare and ice giants are more common \citep{suzuki_exoplanet_2016, Shvartzvald2016, Poleski2021,zang2022,sumi2023}. The inferred population of M dwarf ice-giants is supported by disk theory, which predicts a pile-up of both icy solids and gas in the vicinity of the sublimation radii of volatile species (H$_2$O, CO, N$_2$ etc.) thereby providing a favorable environment for planet formation via core-accretion processes \citep{Okuzumi2016}. Further evidence has been provided by high-resolution imaging of the transitional disk of the low-mass star TW Hya that reveals a gap, a dust/gas ring, and CO tracer species at the predicted ice-line separations \citep{Andrews2016}. In low-mass M dwarf disks, analytical approximations place the primordial CO ice-line at a few 10's of AU \citep{MartinLivio2014} and the ice-lines of other volatile species further out.

Due to high sky backgrounds, ground-based thermal imaging observations of young M dwarfs are only sensitive to planets with masses $\gtrsim1~\mjup$ at separations $\gtrsim$ 10 AU and reveal their occurrence to be $2.3^{+2.9}_{-0.7}\%$ \citep[for masses 2-14 $\mjup$ and separations 8-400 AU]{Lannier2016}. Long term astrometric observations with VLTI/GRAVITY have shown sensitivity to Neptune-mass planets within 5 AU of M stars \citep{gravity_astrometric_2024}, and in the near future, long-baseline high precision astrometric orbits from $Gaia$ are expected to identify thousands of giant exoplanets orbiting M dwarfs \citep{Sozzetti2014}. Recent results from $Gaia$ DR3 are already yielding promising planetary mass candidates, and providing astrometric measurements of some known M dwarf giants \citep{Holl2023,Franson2024, Kiefer2024}. However, the population of sub-Jupiter mass companions orbiting M dwarfs at wide separation ($\gtrsim$10 AU) remains unconstrained and we lack a complete picture of M dwarf planet demographics. On the other hand, the relatively high predicted occurrence rates result in a population of low-mass planets that are both scientifically valuable and potentially detectable using more sensitive direct imaging capabilities.


In addition to their scientifically interesting planet populations, M dwarf stars are ideal candidates for direct imaging surveys for two intrinsic reasons. First, they are the most abundant and long-lived stars in the universe and therefore provide the largest population of nearby target stars for imaging. Second, their low intrinsic brightness results in more favorable star to planet contrasts for direct imaging \citep{beichman_imaging_2010, Brande2020, Carter2021}. The amount of residual starlight not suppressed by the coronagraph is proportional to the apparent brightness of the star, therefore fainter host stars allow fainter (and thus lower mass) planets to be observed. However, as M dwarfs are often too faint to effectively drive ground-based adaptive optics systems, the outskirts of these systems are among the least observed from the ground, resulting in significant uncertainty in exoplanet population statistics and planetary formation models \citep{clanton_synthesizing_2016, bowler_imaging_2016}. 

Having identified the ideal target stars for an exoplanet direct-imaging survey, we now consider the ideal instrument. The Near InfraRed Camera \citep[NIRCam,][]{rieke_performance_2023} on the JWST~\citep{Gardner2023} was predicted to demonstrate a major improvement in exoplanet direct imaging sensitivity \citep{beichman_imaging_2010, Schlieder2015, Carter2021}, especially in the 3-5 $\mu$m wavelength range where the intrinsic thermal emission of young planets is significant \citep{Morley2014,linder_evolutionary_2019}. NIRCam's coronographic mode was exercised both in the commissioning and Early Release Science (ERS) phases of JWST operation to understand the key capabilities. During commissioning, NIRCam successfully observed HD 114174 B, a white dwarf companion to the G5IV-V star HD 114174 A with a flux contrast of $10^{-4}$ and angular separation of 0.5". This observation surpassed the required $5\sigma$ flux contrast sensitivity of $10^{-4}$ at 1.0", and additionally achieved a background limited flux sensitivity of $5 \times 10^{-7}$ at separations greater than 2" \citep{girard_jwstnircam_2022}. The Direct Exoplanet Imaging ERS team \citep{Hinkley2023} completed the first successful direct observation of an exoplanet with JWST: the 14 Myr old super-Jupiter HIP 65426 b with a flux contrast of $4 \times 10^{-6}$ and separation of 1" from its A2V class host star. These observations confirmed that JWST NIRCam's coronagraphic modes exceed the predicted performance capability by up to a factor of 10 in flux contrast sensitivity \citep{carter_jwst_2022}. NIRCam coronagraphy has since been used to continue characterization of known planetary mass companions \citep{Kammerer2024}, aid in the discovery of new planets \citep{Franson2024}, and identify interesting candidates that were previously out of reach \citep{Ygouf2024,Cugno2024}. The demonstrated capabilities of this instrument mode now enable routine direct observations of sub-Jupiter-mass giant planets orbiting young M dwarf stars at wide separations ($\gtrsim 10$ AU) for the first time. 

In this work, we perform the first direct imaging survey with JWST NIRCam coronography to observe a population of stars with no previous observational detection of wide orbit companions, and we provide the first context on the demographics of sub-Jupiters on wide-orbits around M dwarfs. We describe our coronagraphic imaging observations of 9 nearby, young M dwarfs under the Cycle 1 Guaranteed Time Observation (GTO) program 1184\footnote{PI J. Schlieder --~\url{https://www.stsci.edu/jwst/science-execution/program-information?id=1184}; allocated as part of the NIRCam Exoplanets GTO sub-program.}. In Section \ref{sec:Survey_Design} we motivate the target selection and observation strategy for the survey. In Section \ref{sec:Analysis_Methods}, we describe the data reduction techniques used to measure our sensitivity to planetary mass companions and identify point source candidates. In Section \ref{sec:Results}, we summarize the overall performance of JWST NIRCam in the GTO 1184 observations, catalog newly detected sources, and describe several interesting candidates identified in the data. In Section \ref{sec:Discussion}, we compare the survey results to yield estimates, discuss their implications for exoplanet demographics, describe lessons learned, and comment on future work. Finally, we summarize the key findings of this study in Section \ref{sec:Conclusions}.

\begin{table*}
\centering
\begin{tabular}{l|llrrrrc}
\toprule
\textbf{Short}  & \textbf{Standard}& \textbf{Spec.} & \textbf{Distance} & \textbf{Age}& \textbf{W1} & \textbf{W2} & \textbf{Obs.} \\
\textbf{Name} & \textbf{Name} & \textbf{Type}& \textbf{(pc)} & \textbf{(Myr)}& \textbf{(mag)} & \textbf{(mag)} & \textbf{Group \#} \\
\hline
\textbf{AU Mic} & HD 197481 & M1  & $9.9 \pm 0.10$  & $24 \pm 3,^1$ & 4.50 & 4.00 & 1 \\
\textbf{HIP 17695} & G 80-21     & M3  & $16.1 \pm 0.80$ & $149 ^{+51}_{-19},^1$ & 6.81 & 6.68 & 1 \\
\textbf{TYC 5899} & LP 776-25      & M3  & $16.3 \pm 0.40$ & $149 ^{+51}_{-19},^1$ & 6.77 & 6.60 & 1 \\
\textbf{G 7-34} & G 7-34        & M4  & $13.6 \pm 0.20$ & $149 ^{+51}_{-19},^1$ & 8.01 & 7.82 & 1 \\
\textbf{Fomalhaut C} & * alf PsA C   & M4  & $7.6 \pm 0.07$  & $440 \pm 40,^3$ & 6.99 & 6.80    & 2 \\
\textbf{AP Col} &V* AP Col      & M4.5 & $8.4 \pm 0.07$ & $50 ^{+5}_{-10},^2$ & 6.64 & 6.40 & 2 \\
\textbf{2M J0944}  & G 161-71     & M5  & $13.3 \pm 0.15$ & $50 ^{+5}_{-10},^2$ & 7.41 & 7.20 & 2 \\
\textbf{LP 944-20} & LP 944-20     & M9 & $6.4 \pm 0.04$   & $329 \pm 80,^4$ & 9.13 & 8.81 & 3 \\
\textbf{2M J0443}  & 2MASSI     & M9 & $21.1 \pm 0.45$  & $24 \pm 3,^1$ & 10.83 & 10.48 & 3      \\
 & J0443376+000205     &  &  & & &  &      \\
 \toprule
\end{tabular}

\caption{All stars targeted for observation in JWST GTO program 1184 (PI J. Schlieder). Each target star is listed with its spectral type, distance in parsecs,  estimated age in Myr, and observation group number. The observation group numbers show which stars were imaged in back-to-back observations during GTO 1184 to minimize wavefront error (WFE) drift. Superscripts in the age column denote the stellar aging methods used: moving group membership \cite[1,][]{bell_self-consistent_2015} and \cite[2,][]{zuckerman2019}, association with Fomalhaut A \cite[3,][]{mamajek_age_2012}, or lithium depletion \cite[4,][]{pavlenko_lithium_2007}}
\label{table:Targets}
\end{table*}

\section{Survey Design}
\label{sec:Survey_Design}

\subsection{Target Selection} \label{ssec:Target_Selection}

When developing the GTO 1184 survey, we aimed to achieve the deepest companion mass sensitivity limits of any exoplanet direct imaging program and to access new mass-separation parameter space to further understand the wide-orbit M dwarf planet population. Our focus on M dwarfs, with their intrinsically low luminosities, allows deep sensitivity even in the contrast limited regime of NIRCam. We further selected for the youngest and closest of these stars to maximize sensitivity to warm, self-luminous planets and to access the smallest planet-star separations. This combination of intrinsically low luminosity, youth, and proximity made the M dwarfs in nearby young moving groups and associations \citep{Gagne2018} some of the most optimal targets. In 2014 and 2015, we searched through several dozen literature sources reaching back more than a decade to identify all late-K and M dwarf stars confirmed or proposed as members of young moving groups and associations and compile a candidate list \citep[e.g.][]{Zuckerman2004,Torres2008, Shkolnik2012,Malo2014,Kraus2014}. We then supplemented this with a literature search for additional very nearby, young field stars not associated with a known group or association \cite[e.g.][]{mamajek_age_2012, Mamajek2013}. This led to a list of more than 400 candidate targets, which were then vetted for known multiple systems and close background sources in projection using further literature crosschecks and archival imaging data in order to limit known contaminating sources for high contrast imaging \cite[e.g.][]{Worley1984,Wycoff2006,Janson2012,Janson2014}. 

These selection criteria led to a shortlist of more than 50 young M dwarf targets within 25 pc. We further scrubbed this list to only the closest stars and made sub-selections to include targets with a broad range of M spectral types and luminosities. This final part of the selection and prioritization process was subjective by design to include a variety of targets and provide a final data set that enables exploration of NIRCam coronagraphy capabilities in this early survey. We also continued to monitor the literature for new targets or updated information on our prioritized list over the years. This included crosschecks with Gaia data \citep{gaia_collaboration_gaia_2022} and the consideration of newly identified young moving group members and candidates based on those data \citep[e.g.,][]{Gagne2018}. The final target list was solidified in 2020 and consisted of nine of the closest, youngest M dwarfs, as detailed in Table~\ref{table:Targets}.

Here we provide a few notes on the targets. Two, AU Microscopii (AU Mic) and Fomalhaut C, are already known to have extended debris disks \citep{kalas_discovery_2004,kennedy_discovery_2014}. ~\cite{Lawson2023} provides a focused analyses of the AU Mic disk which is detected at 3 - 5 $\mu$m wavelengths for the first time in GTO 1184. \cite{Lawson2024} describes our first-of-its-kind direct-imaging detection of the very faint Fomalhaut C disk, adding a new entry in the short list of M dwarfs with debris disks resolved in reflected light. Each of these papers also reports on the deep sensitivity to wide-orbit planets achieved in those systems, and the independent analyses performed later in this manuscript support those results. AU Mic also has two confirmed transiting planets with semimajor axes 0.065 and 0.11 AU \citep{plavchan_planet_2020,martioli_new_2021}, corresponding to angular separations of 6.6 milli-arcseconds (mas) and 10 mas which are far interior to the NIRCam coronagraph inner working angle (IWA, defined as the separation within which more than 50\% of the incident light is suppressed) and thus undetectable in these observations. TYC 5899 was the latest addition to the target list. It replaced a star that was revealed to potentially be $>$1 Gyr old with the inclusion of new observations and analyses \citep[GJ 393, ][]{Schaefer2018}. TYC 5899 has a known star within a few arcseconds (see Fig.~\ref{fig:Mosaic444}), but we included it as a target due to its youth, spectral type, and proximity. Target 2M J0443's model predicted mass is substellar given its luminosity and young age \citep{Reiners2009,Allers2013}. We include it in the survey to explore NIRCam's contrast performance with a faint target and to search for wide-orbit planets around a substellar target.  

\subsection{Observation Strategy} \label{ssec:Observation_Strategy}

We observe each target with NIRCam for approximately 1 hour in the F444W filter centered on 4.44 $\mu$m (3.881 - 4.982 $\mu$m). This is the longest wavelength, wide-bandpass filter available for use with the NIRCam coronagraphic modes, which maximizes the detection of planet flux for a given exposure time \citep[models predict young giant exoplanets to be bright in the 4-5 $\mu$m wavelength range,][]{linder_evolutionary_2019}. We use the readout pattern SHALLOW2 for the observations of AU Mic with 35 integrations in the F444W filter. We use the MEDIUM8 pattern for every other target with 17 integrations per exposure and 10 groups per integration  as defined in the \href{https://jwst-docs.stsci.edu/jwst-near-infrared-camera/nircam-instrumentation/nircam-detector-overview/nircam-detector-readout-patterns}{NIRCam Detector Readout Patterns page of the JWST User Documentation}\footnote{\url{https://jwst-docs.stsci.edu/jwst-near-infrared-camera/nircam-instrumentation/nircam-detector-overview/nircam-detector-readout-patterns}}. We also observe each target for approximately 30 minutes in the F356W filter centered on 3.56 $\mu$m (3.135 - 3.981 $\mu$m), using the same readout settings except reducing the number of integrations per exposure to 17 for AU Mic, and 8 for the other targets. This enables color-based rejection of background contaminants such as stars and the most common galaxies, as we expect young giant exoplanets to be significantly red in the 3-5 $\mu$m wavelength range \citep{Morley2014, linder_evolutionary_2019}. The total observation time in both filters is split evenly across two roll angles separated by $\sim$10$^{\circ}$, which is typically the maximum allowed by the observatory due to solar avoidance restrictions. The two roll angles enable stellar point-spread function (PSF) subtraction via angular differential imaging \citep[ADI,][]{Marois2006}. ADI leverages the fact that the stellar PSF is fixed to the orientation of the observatory while the position of the planet remains fixed on the sky to disentangle the star and planet flux. All of the aforementioned observations use the subarray SUB320 with a field of view of 20 x 20 arcseconds, enabling the observation of companions at separations up to $\sim$200 AU in GTO 1184, depending on the distance to the individual target star. Finally, each observation uses the coronagraphic mask MASK335R, which is the narrowest round mask available for use with the F444W filter. It provides an inner working angle (IWA) of 0.6'' \citep{Krist2010}, which corresponds to a typical projected separation of $\sim$12 AU in this survey.

Our choice to perform the survey using coronagraphy rather than non-coronagraphic direct imaging was based on detailed predictions for each target using the dedicated software package \code{WebbPSF} for JWST PSF modeling \citep{perrin2012}. We simulated the expected contrast and planet mass sensitivities for each of the 9 targets with MASK335R coronagraph using the F444W filter, and compared with the direct imaging mode using the F480M filter. We explored a range of possible pre-launch wave-front error (WFE) drift values between the science target and reference, adopting $\Delta$WFE $\in$ [0, 2, 5, 10] nm. These simulations revealed that for all targets except the faintest two, 2M J0944 and LP 944-20, the direct imaging mode was significantly more sensitive to WFE drift. This was particularly true at angular separations $\lesssim$2''. Thus, if worse case pre-launch WFE drift predictions were reflected in flight performance, coronagraphy would yield equal or better contrast and mass sensitivity performance at small separations for most targets. For this reason, and to preserve the same observing mode for all targets in the interest of a self referenced survey (see below), we chose to perform the survey entirely in NIRCam's coronagraphic mode.

Even with the vast majority of the target starlight suppressed by NIRCam's coronagraphic optical elements, astrophysical sources near the star (a.k.a. off-axis sources) may still be much fainter than the residual starlight that reaches the detector. To address this, an additional star is typically observed to measure a reference stellar PSF, which can then be scaled to and subtracted from the target star observation to reveal faint, close-in sources. This approach is known as reference differential imaging \citep[RDI,][]{lafreniere_hstnicmos_2009}, and is expected to be more efficient than ADI at removing residual starlight while preserving flux from off-axis sources at small ($\lesssim 2$") angular separations due to the limited range of roll angles available with JWST.

\begin{figure*}
    \centering
    \includegraphics[width=0.8\linewidth]{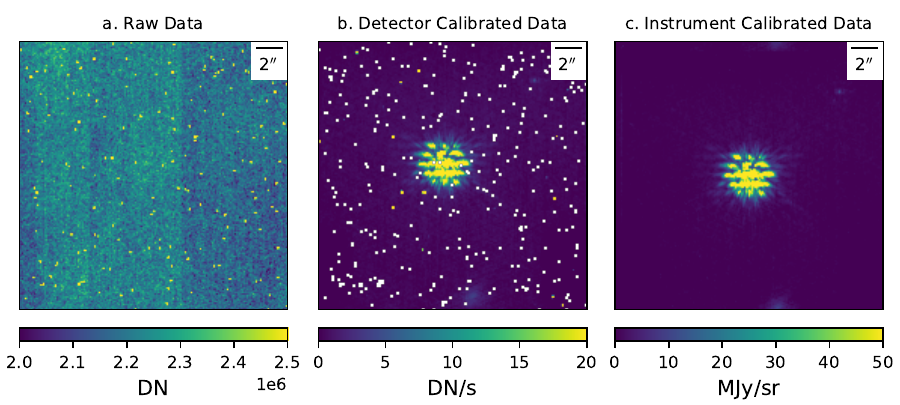}
    \caption{Results of data reduction steps for GTO 1184 target G 7-34 (roll 1 in the F444W filter), using the JWST Calibration Pipeline wrapped by \code{spaceKLIP} \citep{bushouse_jwst_2023,kammerer_performance_2022}. We also include additional custom processing steps for flagging cosmic ray persistence. Panel a shows the raw data as downloaded from the MAST archive. Panel b shows the output of the first stage of the pipeline, which flags and corrects for detector-level errors. Panel c shows the output of the second stage, which corrects instrument-specific errors, smooths over unreliable pixels, and converts the image to physical fluxes. Significant residual starlight is still present in the center of the frame, which needs to be removed with PSF subtraction techniques.} 
    \label{fig:Calibration}
\end{figure*}

In contrast to previous coronagraphic observations with JWST, we do not allocate observing time to dedicated reference stars. We instead use the entire set of science observations as a library of potential reference PSFs to maximize the number of science targets that can be observed within the survey time allocated to GTO 1184. This strategy is made more feasible given that all of the target stars are of similar spectral type, minimizing the variation due to wavelength-dependent PSF structure. We observe the targets in three unbroken sequences grouped by early-, mid-, and late-type M dwarfs to further reduce wavefront error drift between adjacent observations (See Table \ref{table:Targets}). Grouping the targets by spectral type does occasionally result in slews between each target of $\sim$$50^\circ$ or more, however pre-launch thermal and optical modeling (supported later by measurements during commissioning) indicated that the observatory would be stable enough to handle these without introducing significant additional wavefront error \citep{McElwain2023}.

The GTO 1184 observations were executed successfully\footnote{The data is available at MAST: \dataset[doi: 10.17909/1zm1-4x90]{\doi{10.17909/1zm1-4x90}}} between September 6th and November 27th, 2022 (UTC) with the exception of the second roll angle observation of TYC 5899, which failed due to a target acquisition error. Then, the two rolls of TYC 5899 were re-observed in 2023 on different dates due to a sequential observation link error in the observing plan. The two rolls were observed on February 11th and 22nd (UTC), respectively. All three observations of TYC 5899 were used in the analysis, however the high proper motion of the star \citep[243 mas yr$^{-1}$,][]{gaia_collaboration_gaia_2022} and the 4-month delay between the first and last observations result in the apparent blurring of off-axis background sources, so only the final two observations are shown in the survey images and used for off-axis PSF fitting.

\section{Analysis Methods} \label{sec:Analysis_Methods}

The following subsections detail our approach to data reduction, post-processing, and the search for and characterization of point source candidates. 

\subsection{Data Reduction} \label{ssec:Data_Reduction}

The raw JWST data, comprised of non-destructive detector array reads averaged over each group of frames within each integration and exposure (shown in panel a of Figure \ref{fig:Calibration}), becomes accessible on the Mikulski Archive for Space Telescopes (MAST) within about 48 hours of the observations being executed. However, significant calibration and post-processing is required before the data can be used for science analysis and interpretation. The Space Telescope Science Institute (STScI) has thus developed the JWST Science Calibration Pipeline (a.k.a. \code{jwst}) to perform the reduction of all JWST data, composed of three stages described below \citep{bushouse_jwst_2023}.

Stage 1 applies detector-level calibrations to the raw data, including checking for pixel saturation and persistence, as well as performing dark field subtraction, detector gain corrections, and ramp slope fitting. The ramp slope refers to the rate of flux signal build up in each detector pixel during the integration, and it should be linear if the astrophysical source is not variable over the integration timescale. Therefore, if anomalies such as cosmic ray (CR) hits cause significant deviations from the linear ramp slope between two adjacent groups (called ``ramp jumps"), this can be detected and the slope can be fit along each jump-free segment. The threshold for flagging ramp jumps, in number of sigmas above the noise, is one of many calibration parameters which can be specified by the user. The output of this stage is a 3D array of the uncalibrated photon rate in each pixel during each integration, with an example shown in panel b of Figure \ref{fig:Calibration} after averaging over the integrations.

Stage 2 provides physical and instrument-specific corrections to individual integrations to generate a fully calibrated exposure. For the purpose of this study we are concerned only with the NIRCam image calibration steps, which include applying World Coordinate System (WCS) information to transform between pixels and physical coordinates, flat field corrections from instrument and detector response, and flux calibration to convert between counts per second and MJy per steradian. The output of this stage is a 3D array of the calibrated flux (in MJy/str) in each pixel during each integration. The calibrated image for each exposure can then be attained by simply averaging over the integrations, as shown in panel c of Figure \ref{fig:Calibration}.

\begin{figure*}
    \includegraphics[width=\linewidth]{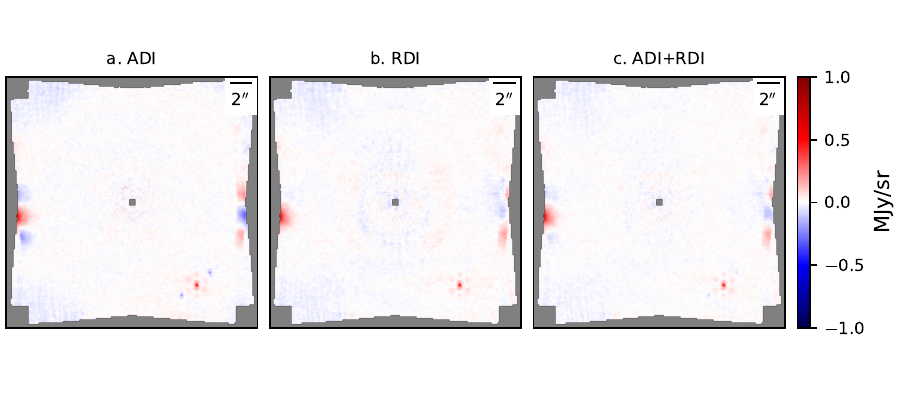}
    \caption{Results of PSF subtraction via ADI, RDI, and ADI+RDI using \code{spaceKLIP}, for G 7-34 in the F444W filter. The positive signal is shown in red, and the negative signal (due to subtraction artifacts) is shown in blue. Panel a shows the ADI result, where each off-axis source has clear negative residuals due to the $10^\circ$ roll angle difference. Panel b shows the RDI result using only HIP 17695 (the target star with the cleanest FOV) as a reference, and panel c shows the ADI + RDI result, which is performed just like the RDI method but including the second roll of the science target as an element in the reference PSF library \citep{kammerer_performance_2022}.} 
    \label{fig:DifferentialImaging}
\end{figure*}

Finally, stage 3 performs additional processing for observational techniques that require the combination of multiple exposures, such as stellar PSF subtraction with ADI and/or RDI. The stage 3 module for coronagraphy applies CR and outlier flags to all science target and PSF reference observations, aligns the PSF references with each target image, and performs the desired PSF subtraction method via Karhunen-Lo\`eve Image Processing \citep[KLIP,][]{soummer_detection_2012}.  

\subsubsection{Pre-Processing}\label{sssec:Pre_Processing}

The ``pre-processing" steps encompass everything leading up to PSF subtraction, including the steps shown in Figure \ref{fig:Calibration}. In addition these include the alignment of images for each science target exposure as well as the preparation of reference PSF images.

Given that the JWST pipeline was developed to cover the broadest possible set of applications, we chose instead to use the publicly available python package \code{spaceKLIP}\footnote{The specific version used in this study is linked in the Acknowledgments} which tunes the 
JWST pipeline specifically for high-contrast coronagraphic imaging of exoplanets \citep{kammerer_performance_2022}. \code{spaceKLIP} essentially wraps the first two JWST pipeline stages in a set of user-friendly python functions and provides custom steps to account for phenomena such as erroneous ramp jump detections, low quality dark current subtraction, flux variations between integrations, and hot or otherwise unreliable pixels. We also use the \code{spaceKLIP} steps for background subtraction, masking and replacing bad pixels, correcting the coronagraph location, and aligning the frames. In general, we adopt the default settings for the step parameters, however we did reduce the jump detection threshold from $4\sigma$ to $3\sigma$, and reduced the sigma clipping threshold in the \code{spaceKLIP} bad pixel cleaning step from 5 to 3, which improved the flagging of cosmic ray hits.

The most significant hurdle we encountered during this phase was the unflagged persistence of several CR hits. The \code{spaceKLIP} ramp-fitting step successfully flags pixels in the first integration affected by a given CR, however for a large enough hit, core pixels can remain warm across multiple integrations in an exposure, such that it can mimic a faint off-axis PSF in the final post-processed image. CR persistence can be unambiguously differentiated from an astrophysical source by inspecting the individual integrations in the affected exposure. As the CR masking by the \code{jwst} pipeline is imperfect, residuals will be clearly visible in the affected frames, and can be masked manually before PSF subtraction. In our first reductions of the survey data, these artifacts led to several spurious ``detections" of a faint source in F444W with no detection in F356W, mimicking the expected signal of a sub-Jupiter mass planet. To ensure that this effect did not cause further issues, we thoroughly examined every group of each observation for CR persistence by subtracting the per-pixel median of the integrations in an exposure from the mean of those integrations and identified places where an outlier existed only in one or a few integrations out of the set. CR persistence stands out using this method and we flagged each instance by eye and masked those pixels only in the affected integrations before higher level processing.

\subsubsection{KLIP Processing}\label{sssec:Post_Processing}

PSF subtraction is performed via the  Karhunen-Lo\`eve Image Processing \citep[KLIP,][]{soummer_detection_2012}, which uses a Karhunen-Lo\`eve transform of the reference observations to build an orthogonal basis of eigenimages (a.k.a.~KLIP modes) that are then fit to the science observation. The \code{spaceKLIP} implementation of this method is based on the python package \code{pyKLIP} \citep{wang_pyklip_2015} and can perform PSF subtraction via ADI and RDI with more flexibility than the \code{jwst} pipeline, as well as a combination of the techniques (ADI+RDI). The science image can also be separated into a number of annuli (concentric circular regions) and position-angle (PA) subsections, so that the PSF subtraction can be performed independently in multiple subsections of the image. This is particularly helpful to avoid bias due to a bright, off-axis source in the science image affecting the PSF subtraction quality in the rest of the image. In our reductions, we use four evenly spaced annuli and no position angle divisions, which provided the best balance of isolating bright off-axis sources while keeping each individual region large enough to avoid error inflation from small sample statistics near the IWA. One can also tune the number of KL modes (i.e.~principal components) used for the final PSF model. We opt to use 6 KL modes to reduce the probability of astrophysical signals being subtracted by the KLIP algorithm, following the prescription used in \citet{carter_jwst_2022}.

ADI is performed by reducing each roll angle of a given science target using the alternate roll angle of that same science target as the reference observation. After each roll angle is PSF-subtracted via the KLIP algorithm as described above, the resulting images are de-rotated and averaged. The presence of the same off-axis sources in the science and reference observations for this method creates a characteristic negative-positive-negative signal in the final image (as in Panel a of Figure \ref{fig:DifferentialImaging}), where the negative artifacts are separated from the positive signal by the same angle that separates the two observational rolls. However at small separations from the host star, this can result in ``self-subtraction,'' where the negative and positive lobes are separated by a distance less than the width of the PSF, thus reducing the throughput of the KLIP algorithm in that region.

RDI is performed by processing each roll angle of a given science target using the other targets as ``references'', before de-rotating and averaging the two rolls.  This method, shown in Figure \ref{fig:DifferentialImaging} Panel b, avoids the self-subtraction effect, however the PSF subtraction quality is more sensitive to spectral type and WFE differences between the science and reference targets as well as to off-axis sources present in the reference observations. 

Finally, ADI+RDI is performed by applying RDI to each roll observation, including the alternate rolls of the science target as well as the other targets in the PSF reference library for subtraction, and then de-rotating and summing the two PSF-subtracted rolls. This enables the use of the same star as a reference while somewhat suppressing the effect of self-subtraction, as shown in Panel c of Figure \ref{fig:DifferentialImaging}.

\subsubsection{PSF Library Preparation} \label{sssec:psf_library}

As this survey does not contain dedicated reference stars, we perform an additional processing step for each science image so that it can also serve as a reference PSF free of contaminating off-axis sources. We first mask each off-axis source revealed in a first-pass ADI reduction of the science targets as described above, using a signal-to-noise ratio per resolution element (SNRE) threshold of 5, then replace each masked pixel in each science frame with the median value of that pixel across all frames in the library. This effectively removes contaminating sources from the reference PSF library while preserving the stellar PSF structure and avoids biasing the subtraction with a uniform fill value. The resulting images are treated as the reference PSF library for the survey. We exclude from the reference library TYC 5899, which had an extremely bright source at about 3" separation, as well as Fomalhaut C and AU Microscopii which host circumstellar disks, making the stellar PSF difficult to sample. The remaining 6 stars are referred to throughout this paper as the ``full reference library''.

\subsection{Image Analysis} \label{ssec:Image_Analysis}

\begin{figure}
    \includegraphics[width=\linewidth]{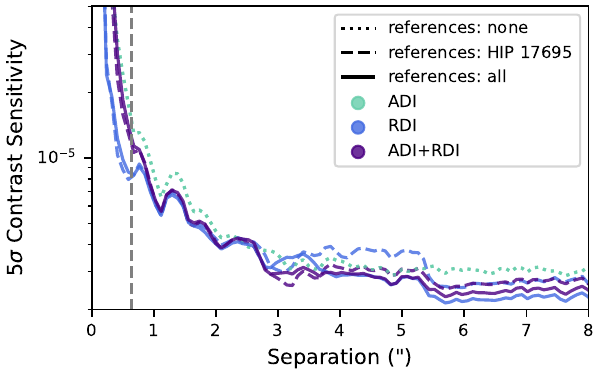}
    \caption{5$\sigma$ contrast sensitivity curves for different PSF subtraction techniques and reference PSF choices applied to the G 7-34 observation in F444W, as a function of angular separation from the target star. The teal, blue, and purple lines show the results for ADI, RDI, and ADI+RDI respectively. Reductions using the full library of PSF references are shown with a solid line, those using only HIP 17695 as a reference are shown with a dashed line, and those using no references (ADI only) are shown with a dotted line. The 0.6'' coronagraph inner working angle (IWA) is shown as a vertical gray dashed line. We can see that RDI and ADI+RDI outperform ADI alone at nearly all separations, regardless of the number of references used.}
    \label{fig:SubtractionContrast}
\end{figure}

Because of the artifacts produced during PSF subtraction, especially in the case of ADI, photometric analysis cannot be performed through traditional aperture photometry. Instead we use the \code{spaceKLIP} routines developed to measure contrast sensitivity and off-axis source photometry via PSF forward modeling and injection-recovery. 

We characterize the sensitivity of our observations based on the 5-$\sigma$ flux contrast sensitivity, (a.k.a. the ``$5\sigma$ contrast") which the is the flux contrast of an off-axis source relative to the star that could be detected with a signal-to-noise ratio of 5, equivalent to a false alarm probability of $2.9 \times 10^{-7}$. This is determined by sampling the radial noise profile at twice the angular resolution of the observing system and quantifying the $5\sigma$ flux threshold using Student t-statistics.  The combined throughput of the PSF subtraction algorithm is then accounted for by injecting an off-axis PSF model with a pre-determined flux \citep[generated by \code{WebbPSF},][]{perrin2012} into the pre-PSF-subtracted images on a grid of planet-star separations between 0.1" and 5" and position angles separated by $60^\circ$. We then perform the PSF subtraction, recover the resulting source photometry by fitting a 2D gaussian via least squares optimization, and divide by the input flux to measure the dimensionless throughput as a function of separation. This is implemented in a routine provided by \code{pyKLIP} and wrapped by \code{spaceKLIP}, which includes the effects of small-sample statistics at narrow separations following \citet{mawet_fundamental_2014}. The throughput of the NIRCam coronagraphic optical system is then accounted for by applying the transmission map for the M335R mask configuration provided by \code{WebbPSF}.

We used the G 7-34 system as an example to study the effect of different PSF subtraction methods and reference PSF libraries on the final contrast sensitivity. Figure \ref{fig:SubtractionContrast} shows the results of PSF subtraction of using ADI with no reference stars, and using RDI or ADI+RDI combined with either HIP 17695 as a reference, or with all of the good quality science observations as references. In this example, the inclusion of additional reference stars improves contrast sensitivity in the background-limited regime beyond $\sim 3$ arcseconds. However within the speckle-limited regime of $\lesssim 3$ arcsec, we do not see the same improvement. We understand this to be caused by G 7-34 and HIP 17695 being very closely matched both in wavefront error (due to the back-to-back observations) and in star-coronagraph alignment (by coincidence). The mean difference in star-coronagraph alignment between the G 7-34 observations and the HIP 17695 observations is 0.088 pixels, compared to 0.155 pixels between G 7-34 and the rest of the library. Thus, as we add more reference observations, which have more significant differences in wavefront error and coronagraph alignment compared to HIP 17695, we do not see significant improvements in the PSF subtraction quality within 3 arcsec. 

We note that RDI with the full reference library results in the deepest contrast sensitivity both within the IWA, denoted by the vertical dashed line in Fig.~\ref{fig:SubtractionContrast}, and in the background-limited regime beyond 5 arcsec, therefore we use this method for our primary analysis. We do however cross check with the alternate methods to ensure that a given off-axis source is astrophysical and not an artifact of one of the subtraction algorithms.

\begin{figure}
    \centering
    \includegraphics[width=\linewidth]{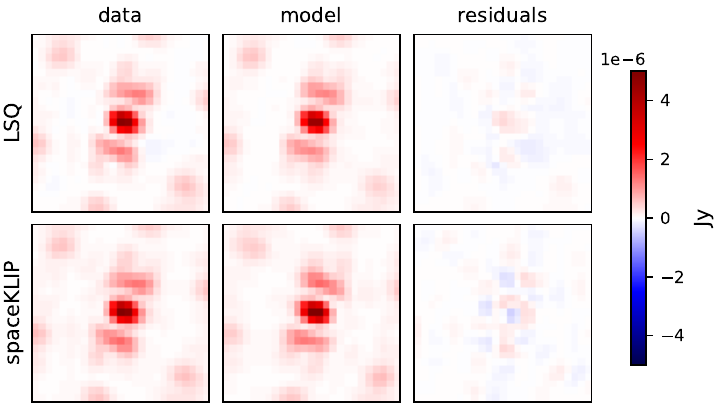}
    \caption{Examples of the two different approaches used for forward-model PSF fitting, showing source P1 near TYC 5899 as an example. The left column shows the image data, the center column shows the best-fit PSF model, and the right column shows the residuals. The result from the least-squares PSF fitting routine are shown in the top row, while the default spaceKLIP fitting routine is shown in the bottom row and produces a similar result.} 
    \label{fig:psf_fit}
\end{figure}

The \code{spaceKLIP} routine for measuring the photometry of an off-axis source, similar to the contrast curve measurement, generates an off-axis PSF model at the approximate planet-star separation of the companion, then applies a forward model of the stellar PSF subtraction method used in the data reduction, and finally fits the resulting off-axis PSF model to the data using a Markov Chain Monte Carlo (MCMC) approach. This allows us to measure the luminosity of the off-axis source relative to the target star while accounting for any self-subtraction or local under/over-subtraction due to the KLIP algorithm. An example of this method applied to a well-detected, off-axis source in the vicinity of TYC 5899 is shown in the bottom row of Figure \ref{fig:psf_fit}.

In some cases with lower SNR or significantly non-zero background, the \code{spaceKLIP} photometry measurement produces a result that is visually erroneous in diagnostic images, so we have developed a semi-independent routine to measure the photometry of these objects. We produce a normalized PSF model using \code{WebbPSF}, which we fit to the data using a least-squares (LSQ) minimization. We then use the separation-dependent measurement of the KLIP algorithm and coronagraph mask throughput measured by \code{spaceKLIP} during the contrast curve calculation to convert from the PSF model amplitude to the source flux. We determine the measurement uncertainty via the bootstrap method, wherein the model fit residuals are shuffled in the image plane, the shuffled residuals are added back to the data, and the model fit is repeated 1000 times to produce a posterior distribution of best-fit parameter values. This produces a similar result (typically agreeing within 1-2 $\sigma$) to the \code{spaceKLIP}-derived photometry and astrometry for robustly detected sources, as shown visually in  the top row of Figure \ref{fig:psf_fit}. More marginal sources (i.e. SNR $<$ 10) show greater discrepancies between the spaceKLIP and LSQ fit, and in these cases we report higher astrometric uncertainties to account for this.

\section{Results} \label{sec:Results}

\begin{figure*}[p]
    \includegraphics[width=\linewidth]{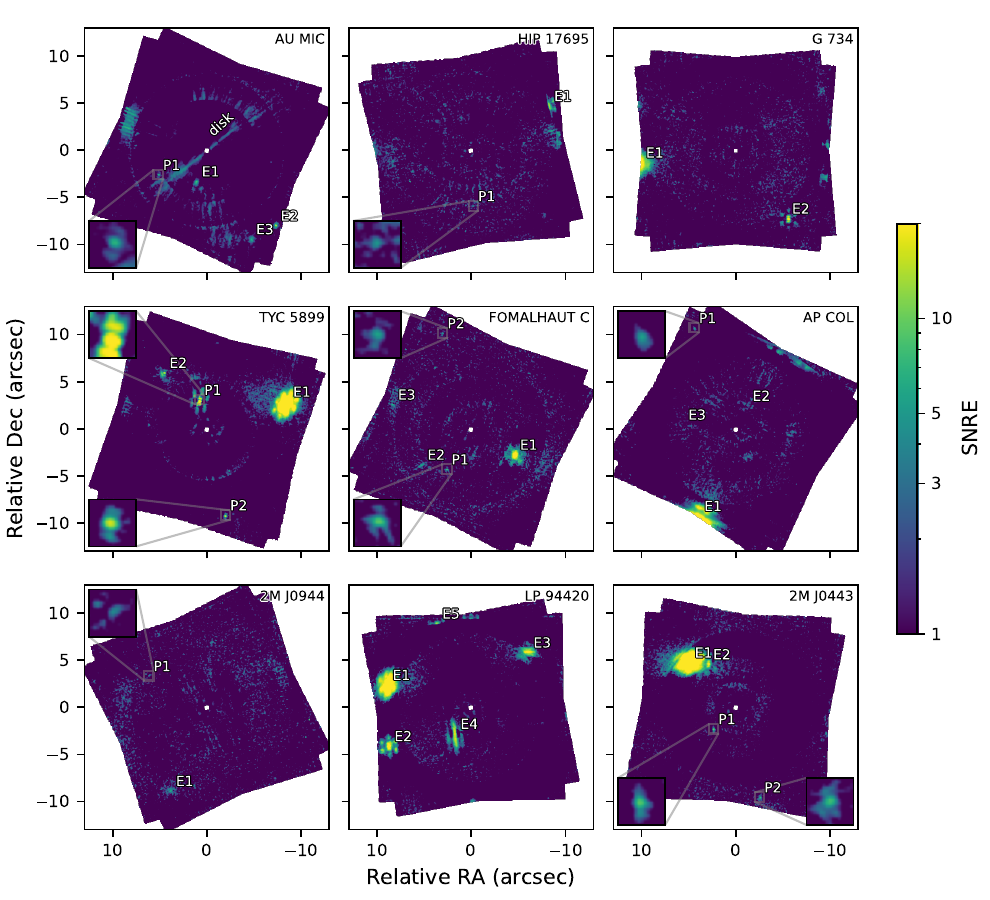}
    \caption{Signal-to-noise ratio per resolution element (SNRE) maps of all targets in GTO 1184, in the F444W filter. All observations were reduced with RDI using the full library of reference PSFs. Point-like and extended sources are labeled in the images (with names beginning with ``P" and  ``E" respectively), and the point-like sources are magnified in the image insets. Extended background sources are abundant in the survey, and the edge-on disk of AU Mic is clearly detected.}
    \label{fig:Mosaic444}
    
\end{figure*}

\begin{figure*}[p]
    
    \includegraphics[width=\linewidth]{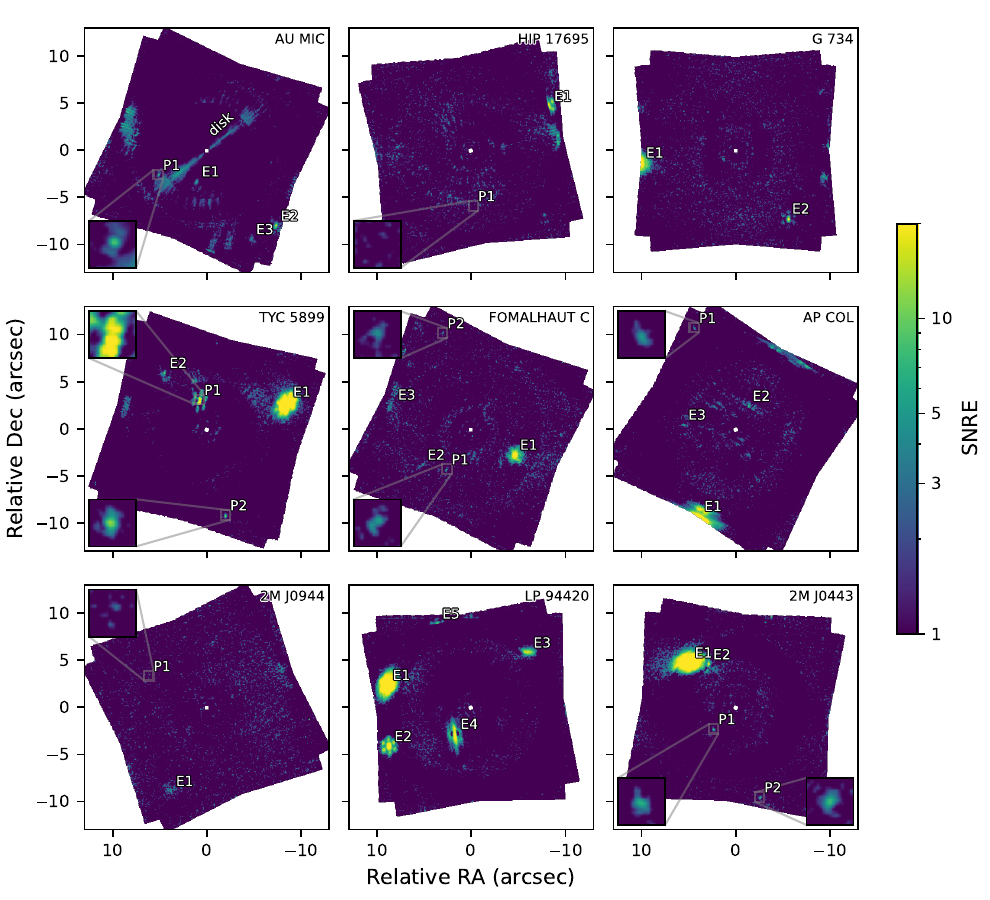}
    \caption{Signal-to-noise ratio per resolution element (SNRE) maps of all targets in GTO 1184, in the F356W filter. All observations were reduced with RDI using the full library of reference PSFs. Point-like and extended sources are labeled in the images (with names beginning with ``P" and  ``E" respectively), and the point-like sources are magnified in the image insets. In cases where the candidate source is not detected in the F356W filter, we still label it and provide an inset image to demonstrate its non-detection. Extended background sources are abundant in the survey, and the edge-on disk of AU Mic is clearly detected.}
    \label{fig:Mosaic356}
    
\end{figure*}

\begin{sidewaystable}
\centering
\begin{tabular}{llrr|rrr|rrr|c|l}
\toprule
 &   &   & \multicolumn{1}{c|}{}   & \multicolumn{3}{c|}{\textbf{F444W}} & \multicolumn{3}{c|}{\textbf{F356W}} & \textbf{F356W-F444W}  & \\
\textbf{Target} & \textbf{Source} & \textbf{RA} & \textbf{Dec} & \textbf{SNRE} & \textbf{Flux} & \textbf{SNR} & \textbf{SNRE} & \textbf{Flux} & \textbf{SNR}  & \textbf{Color} & \textbf{Method} \\
        &        & ["] & ["] &    & [mag] &  \multicolumn{1}{c|}{}  &   &   [mag] &   \multicolumn{1}{c|}{}    & [mag] &  \\
\hline
AU Mic & P1 &      $5.057^{+0.014}_{-0.014}$ & $-2.510^{+0.011}_{-0.011}$ &  7.21 & $18.947^{+0.144}_{-0.127}$ &  7.68 &  9.26 & $19.203^{+0.118}_{-0.106}$ &  7.89 & $0.257^{+0.186}_{-0.166}$ & sklip \\
TYC 5899 & P1 &    $0.745^{+0.007}_{-0.007}$ &  $3.031^{+0.007}_{-0.007}$ & 68.69 & $14.634^{+0.004}_{-0.004}$ & 75.11 & 68.46 & $14.804^{+0.006}_{-0.006}$ & 68.02 & $0.170^{+0.007}_{-0.007}$ & sklip \\
TYC 5899 & P2 &   $-1.979^{+0.008}_{-0.008}$ & $-9.107^{+0.007}_{-0.007}$ & 12.97 & $19.540^{+0.032}_{-0.032}$ & 33.86 & 11.38 & $20.003^{+0.041}_{-0.041}$ & 25.63 & $0.463^{+0.052}_{-0.052}$ & lsq \\
Fomalhaut C & P1 & $2.521^{+0.058}_{-0.058}$ & $-4.219^{+0.112}_{-0.112}$ &  9.03 & $20.980^{+0.174}_{-0.150}$ &  6.83 &  5.41 & $21.688^{+0.111}_{-0.101}$ &  5.84 & $0.708^{+0.207}_{-0.181}$ & sklip \\
Fomalhaut C & P2 & $2.992^{+0.042}_{-0.042}$ & $10.216^{+0.014}_{-0.014}$ &  5.76 & $21.153^{+0.108}_{-0.098}$ &  5.76 &  4.59 & $21.997^{+0.157}_{-0.138}$ &  4.10 & $0.844^{+0.191}_{-0.169}$ & sklip \\
 AP Col & P1 &      $4.385^{+0.037}_{-0.037}$ & $10.692^{+0.013}_{-0.013}$ &  5.81 & $21.042^{+0.125}_{-0.112}$ &  4.04 &  5.39 & $21.828^{+0.215}_{-0.180}$ & 3.72 & $0.786^{+0.249}_{-0.212}$ & sklip \\
2M J0443 & P1 &    $2.311^{+0.011}_{-0.010}$ & $-2.259^{+0.011}_{-0.010}$ &  5.94 & $20.678^{+0.085}_{-0.068}$ & 11.29 &  5.84 & $21.334^{+0.142}_{-0.086}$ &  9.44 & $0.657^{+0.165}_{-0.109}$ & lsq \\
2M J0443 & P2 &   $-2.537^{+0.056}_{-0.056}$ & $-9.475^{+0.009}_{-0.009}$ &  8.44 & $20.865^{+0.095}_{-0.088}$ &  5.96 &  7.99 & $21.444^{+0.083}_{-0.077}$ &  7.09 & $0.579^{+0.126}_{-0.117}$ & sklip \\

\toprule
\end{tabular}

\caption{We detected 8 point-like sources with $\ge$5$\sigma$ significance in both F444W and F356W. We report each source (as labeled in Figures \ref{fig:Mosaic444} and \ref{fig:Mosaic356}) with its position relative to the target star; as well as the peak SNRE, apparent magnitude from PSF fitting, and PSF fit SNR in each filter; the two-filter color, and the method used for PSF fitting as described in Section \ref{ssec:Image_Analysis}. As we expect sub-stellar companions to have a F356W-F444W color greater than $\sim 2$, none of these sources have been flagged as planetary mass companion candidates (See Figure \ref{fig:isochrones}).}
    
\label{table:robust_photom}
\end{sidewaystable}

\begin{sidewaystable}
\centering
\begin{tabular}{llrrc|rrr|rc|c|l}
\toprule
 &   &   &  & & \multicolumn{3}{c|}{\textbf{F444W}} & \multicolumn{2}{c|}{\textbf{F356W}} & \textbf{F356-F444W} & \\
\textbf{Target} & \textbf{Source} & \textbf{RA} & \textbf{Dec} & \textbf{Proj. Sep.} & \textbf{SNRE} & \textbf{Flux} & \textbf{SNR} & \textbf{SNRE} & \textbf{3$\sigma$ Flux Limit} & \textbf{3$\sigma$ (1$\sigma$)} \textbf{Color} & \textbf{Method} \\
 &  & ["] & ["] & [AU] &  & [mag] & \multicolumn{1}{c|}{} &  & [mag] & [mag] &  \\
\hline
HIP 17695 & P1 & $-0.281^{+0.007}_{-0.007}$ & $-5.837^{+0.009}_{-0.007}$ & 94.1 & 4.84 & $22.481^{+0.282}_{-0.195}$ & 1.64 & -0.389 & $>22.865$ & $>$0.38 ($>$1.58) & lsq \\
2M J0944 & P1 &   $6.098^{+0.018}_{-0.018}$ &  $3.423^{+0.022}_{-0.022}$ & 93.0 & 4.02 & $21.842^{+0.093}_{-0.085}$ & 3.91 &  2.976 & $>22.863$ & $>$1.02 ($>$2.21) & sklip \\
\toprule
\end{tabular}
\tablecomments{In regions where the background or noise fluctuation is slightly negative and there is no significant astrophysical flux, the SNRE can be slightly negative.}

\caption{We detected 2 point-like dropout sources with marginal (SNRE $>$4$\sigma$) flux in F444W and insignificant (SNRE $<$3$\sigma$) flux in F356W. We report each source (as labeled in Figures \ref{fig:Mosaic444} and \ref{fig:Mosaic356}) with its position relative to the target star, as well as the  projected separation in AU. We also show the peak SNRE in F444W, the apparent magnitude and SNR from PSF fitting in F444W, the SNRE and 3$\sigma$ magnitude limit from the contrast curve at the associated location in F356W, and the corresponding 3$\sigma$ two-filter color limit (the 1$\sigma$ limit is shown in parentheses). The last column shows the method used for PSF fitting as described in Section \ref{ssec:Image_Analysis}. The low PSF fit SNR of HIP 17695 P1 in F444W makes it less compelling as an exoplanet candidate.}
    
\label{table:dropout_photom}
\end{sidewaystable}

We reduced all observations for each target using the RDI technique described in Section \ref{sssec:Post_Processing} with the full PSF reference library. We then calculated the signal-to-noise ratio per resolution element (SNRE) in each pixel by convolving the final image with a 2D full-width-half-max top-hat model (of width 0.114" for F356W and 0.140" for F444W) and dividing by the finite-element-corrected noise profile. The resulting SNRE maps for each target in F444W and F356W are shown in Figures \ref{fig:Mosaic444} and \ref{fig:Mosaic356} respectively, with candidate point sources magnified in the image insets and the extended sources labeled but not magnified. The survey images in physical units (MJy/steradian) are available for reference in Appendix \ref{a:flux_maps}. As noted previously, we do not include the first of the three TYC 5899 observations in these maps as the 4-month delay between the first and second observations, combined with the high proper motion of the target star, causes a non-trivial blurring of background objects.

We recover the bright, edge-on, circumstellar disk of AU Mic, whose analysis is detailed in \citet{Lawson2023}. The circumstellar disk of Fomalhaut C is not obvious in these reductions, however this is because our analysis is optimized to recover point-like sources, rather than diffuse emission. The disk is however detected and analyzed using Model-Constrained RDI in \citet{Lawson2024}. As we aim in this paper to detect and analyze potential substellar companions, we use the best-fit disk models from the papers described above to subtract the disk emission from AU Microscopii and Fomalhaut C before characterizing the contrast sensitivity and photometry of these observations.

\subsection{Survey Sensitivity} \label{ssec:Survey_Sensitivity}

\begin{figure}[h]
    \includegraphics[width=\linewidth]{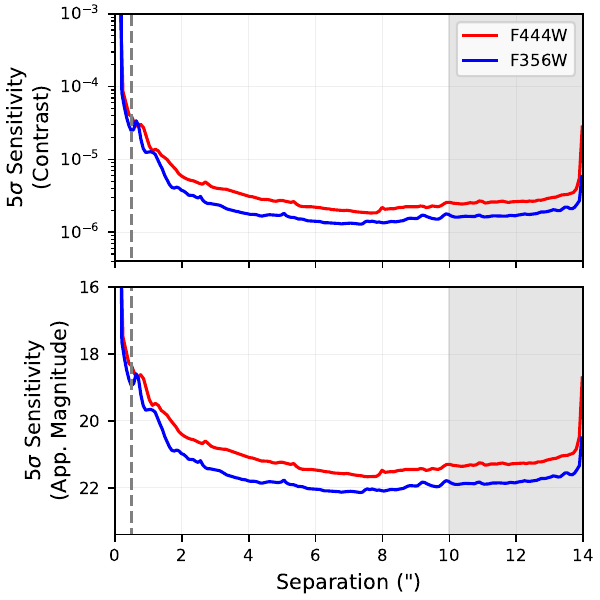}
    \caption{The median $5\sigma$ sensitivity curves for the entire survey in the F444W filter (red lines) and F356W filter (blue lines). The sensitivity is given in units of planet-star flux contrast in the upper panel, and apparent magnitude in the lower panel. The coronagraph inner working angle (IWA) is shown in the gray dashed line. The separations where only partial coverage exists due to square framed observations at multiple roll angles is shown by the gray shaded region. We demonstrate deep limits in absolute flux at close angular separations.}
    \label{fig:SurveyContrast} 
\end{figure}

\begin{figure}
    \includegraphics[width=\linewidth]{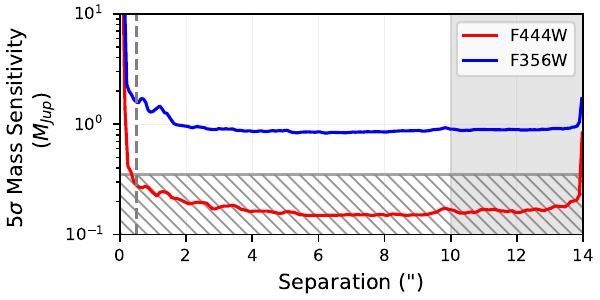}
    \caption{ The $5\sigma$ mass sensitivity curves for the G 7-34 observations, as a function of angular separation from the target star, demonstrate extremely deep mass sensitivity with the F444W filter (in red) as compared to the F356W filter (in blue). Due to this difference in sensitivity, we expect significantly sub-Jupiter-mass companions to appear as an ``F356W dropout'', e.g. being detectable at F444W, but not at F356W. The masses are derived from the \code{BEX-petitCODE} cooling curve models, and the hatched region shows where we have extrapolated the model grid to temperatures below the lower limit of 150 K.}
    \label{fig:ContrastG734}
\end{figure}

\begin{figure*}
    \includegraphics[width=\linewidth]{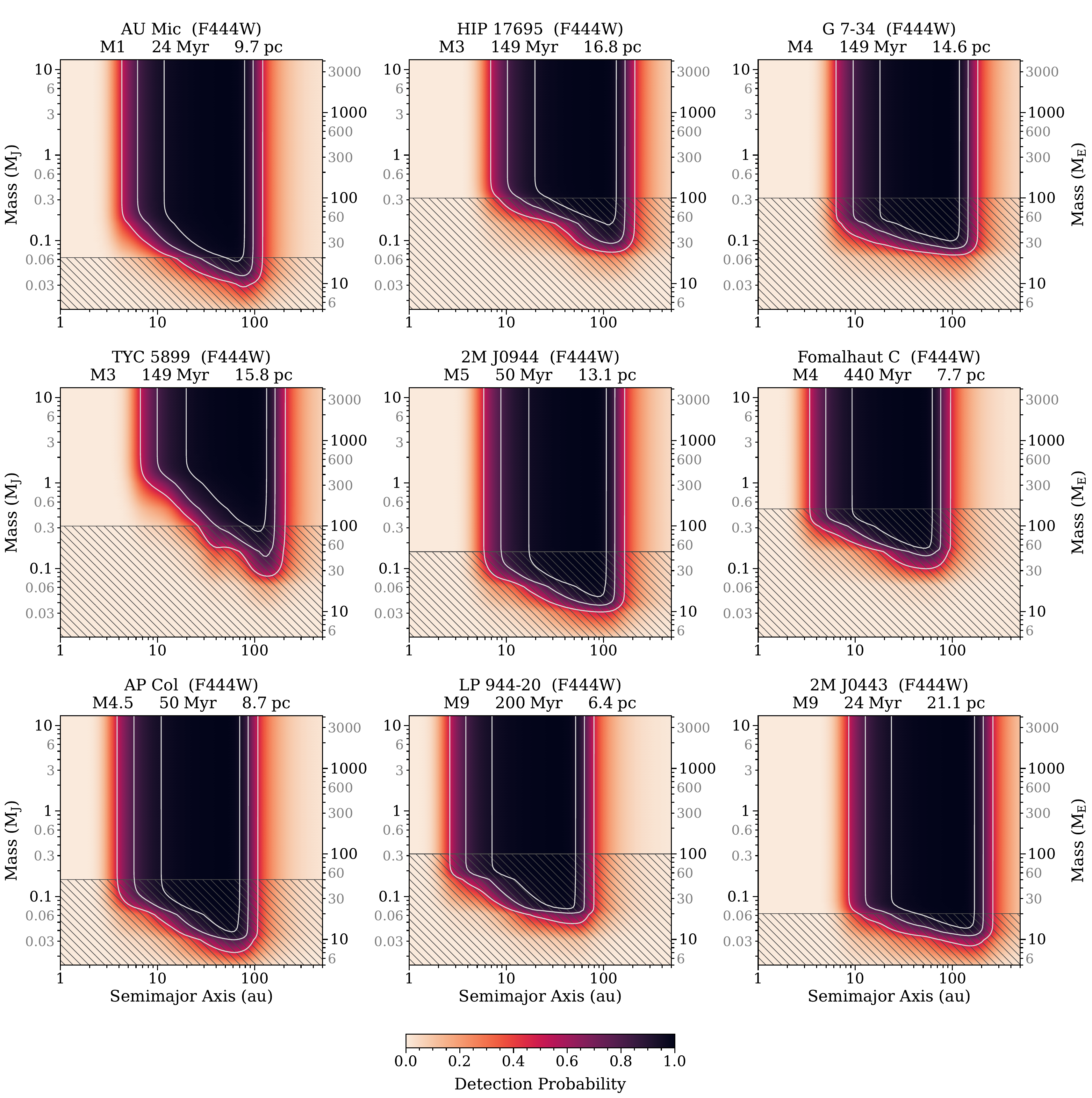}
    \caption{Detection probability maps in mass and semimajor axis space for each survey target, demonstrating sensitivity to Saturn-mass exoplanets around all targets, and specifically to Saturn-mass exoplanets with a Saturn-like semimajor axes (9.5 AU) around at least 6 of the 9 targets. We show the mass sensitivity achieved in the F444W filter, as the \code{Ames-Cond}, \code{ATMOS-CEQ}, and \code{BEX-petitCODE} cooling curves predict sub-stellar objects to be the brightest in this filter as compared to F356W. The solid white lines show detection probability contours of 50\%, 80\%, and 95\%, and the hatched region shows where our observations are sensitive to masses below the lower limit of the available model grids for a given target's age, and thus we have extrapolated the cooling curves to lower masses (lower temperatures) for visualization purposes. The mass sensitivity of the GTO 1184 survey surpasses the low-mass limit of available planet evolution and atmosphere grids for every target, regardless of age or intrinsic brightness.}
    \label{fig:combined_detection}
\end{figure*}

Figure \ref{fig:SurveyContrast} shows the median survey sensitivity in units of flux contrast (upper panel) and apparent magnitude (lower panel) in both filters (F444W in red and F356W in blue). We achieve a median 5$\sigma$ sensitivity of $1.5 \times 10^{-5}$ (and median magnitude sensitivity of 19.3 mag) at a separation of 1" and $2.7 \times 10^{-6}$ (21.2 mag) in the background-limited regime (i.e. at separations $\gtrsim 3$") with the F444W filter, as measured via the methods described in Section \ref{ssec:Image_Analysis}. In the F356W filter, the median sensitivity is $1.3 \times 10^{-5}$ (19.7 mag) at 1" and $1.7 \times 10^{-6}$ (21.8 mag) at 5". For reference, we provide the flux contrast and apparent magnitude sensitivity curves for each target individually in Appendix \ref{a:contrast}.

We then convert the flux contrasts to planet mass sensitivities using published evolution models for stellar- and planetary-mass objects. For masses below 2 $\mjup$ we use~\code{BEX-petitCODE} models \citep{linder_evolutionary_2019}, for masses below 0.075 $\msun$ ($\sim 80~\mjup$) we use \code{ATMOS-CEQ} models \citep{phillips_new_2020}, and for all stellar masses we use \code{Ames-COND} models \citep{allard_2001}. As an example, we show the $5\sigma$ mass sensitivity curves in units of companion mass for target G 7-34 in Figure \ref{fig:ContrastG734}, demonstrating sensitivity to sub-Saturn mass companions in the F444W filter. Despite slightly deeper flux contrasts, F356W has significantly shallower mass sensitivity. This is due both to the intrinsic redness of young planetary mass objects and the shorter exposure time in F356W compared to F444W. Due to this difference in sensitivity, we expect significantly sub-Jupiter-mass companions to appear as ``F356W dropouts'', e.g. being detectable in the F444W observations, but not in F356W. 

In this analysis, we also assume an underlying distribution of orbital eccentricities (Gaussian with a mean of 0 and sigma of 0.3), and uniformly random orientations in space (corresponding to uniform $\cos{i}$, longitude of ascending node [0-$2\pi$], and longitude of periastron in [0-$2\pi$]) to account for projection effects. We can then measure the detection probability as a function of mass and semimajor axis by injecting a population of planets with the above distribution and calculating the percentage of planets that would be detectable with $5\sigma$ confidence based on the apparent magnitude sensitivity curve for each target star. Figure \ref{fig:combined_detection} shows the detection probability in the F444W filter (the more mass-sensitive of the two) in mass and semi-major axis space for each target, with the hatched region denoting where our observations are sensitive to planet temperatures below the 150 K lower limit of the \code{BEX-petitCODE} model grid. In this region we have extrapolated the cooling curves in log space to be able to visualize the probable detection space, though we note that inaccuracies are likely beyond the edge of the model grid. This being said, the extrapolations are not needed to show that we clearly demonstrate sensitivity to Saturn-mass (0.3 $\mjup$) objects in the vicinity of 8 of our 9 targets, and very likely in the case of Fomalhaut C as well. In addition, we are sensitive ($>$50\% detection probability) to Saturn-mass exoplanets at Saturn-like separations (9.5 AU) for at least 6 targets. These are the deepest mass sensitivity limits for any direct exoplanet imaging survey to date, demonstrating the power of JWST coronagraphy to explore the outskirts of young M dwarf systems.

\subsection{Source Catalog} \label{ssec:Source_Catalog}

We use a two-tiered search for off-axis point sources to catalog both robust and marginal detections. For robust detections we use an SNRE threshold of 5.0 in either F444W or F356W, and for marginal dropout detections we require an SNRE greater than 4.0 in F444W and less than 3.0 in F356W. This is due to the predicted relative brightness of sub-Jupiter-mass objects in F444W as compared to F356W. We categorize objects as point-like, as opposed to extended, by visual inspection of the PSF fit residuals. As the PSF fit assumes a point-like flux distribution, extended sources will result in a positive “halo” or other distributions of residual flux around the PSF core. 

Across all targets in the survey, we identify 10 point-like sources. Eight of these are robustly detected in both filters, and we display the measured SNRE, relative astrometry, F444W and F356W magnitudes, and F444W-F356W color, for each source in Table \ref{table:robust_photom}. We also note which PSF fitting method was used for each source, however we report the results of both methods (described in Section \ref{ssec:Image_Analysis}) in an extended table in Appendix \ref{sec:FullPhotometry}.

\begin{figure}[h]
    \centering
    \includegraphics[width=\linewidth]{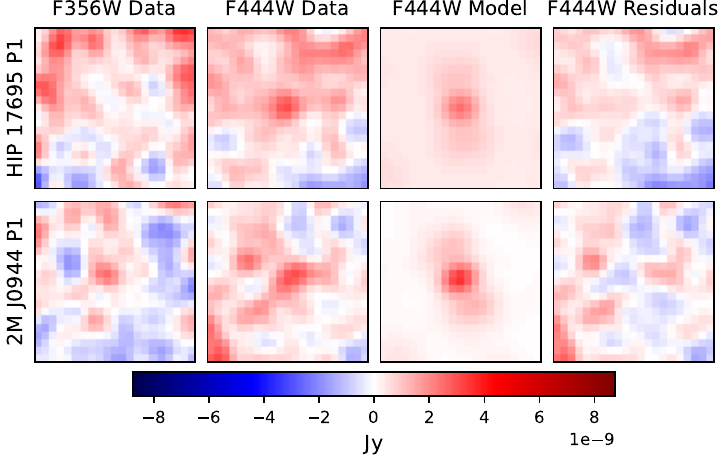}
    \caption{Data and PSF fit results for the marginally detected ``F356W dropout" sources. The first column on the left shows the F356W data, the second column shows the F444W data, the third column shows the best-fit PSF in F444W (using the least-squares fit for HIP 17605 P1 and the spaceKLIP fit for 2M J0944 P1), and the right column shows the residuals in F444W. The PSF fit SNR of HIP 17695 P1 in F444W is 1.64, leading it to be less compelling as an exoplanet candidate. 2M J0944 P1 is detected with an SNR of 3.91, and remains as a marginal detection. As the SNRE of each signal in F356W falls below the threshold of 3, we do not perform a PSF model fit for the data in that filter. We note that the positive blob visible in the F356W data for 2M J0944 P1 is offset by several pixels from the location of the source in F444W, and that it has an amplitude similar to the F444W residuals.} 
    \label{fig:dropout_psfs}
\end{figure}

The remaining 2 sources are marginally detected in F444W (SNRE $>4$), but not in F356W (SNRE $< 3$, a.k.a ``F356W dropouts"). For these sources, we report the SNRE in each filter, relative astrometry, projected separation in AU, F444W magnitude, F356W magnitude limit, and the 3- and 1-$\sigma$ lower limit on the F444W-F356W color (as determined from the contrast sensitivity curve for that target at the same angular separation) in Table \ref{table:dropout_photom}. The data, best-fit PSF model, and residuals for each dropout source in F444W are shown in Figure \ref{fig:dropout_psfs}. In this figure we also show the F356W data at the same position, however as the SNRE of each signal in F356W falls below the threshold of 3, we do not perform a PSF model fit for the data in that filter. While the reader may observe a positive blob in the F356W data for 2M J0944 P1, we note that this blob is offset from the position of 2M J0944 P1 in F444W by several pixels, and is of similar amplitude to the background noise fluctuations that appear in F444W.

As our primary goal is to detect candidate substellar companions, which we expect to be point sources, we do not characterize the extended sources in detail. However we did detect 22 extended sources with SNRE $>5$ in either F444W or F356W, and we catalog their approximate position and peak SNRE in Appendix \ref{sec:ExtendedSources}. Due to an artifact of the \code{spaceKLIP} image alignment step, some of the extended sources near the edge of the field of view are wrapped to the other side of the image. We have noted in the table which sources this affects.

\subsubsection{Background Source Rejection} \label{Background_Source}

\begin{figure*}
    \includegraphics[width=0.33\linewidth]{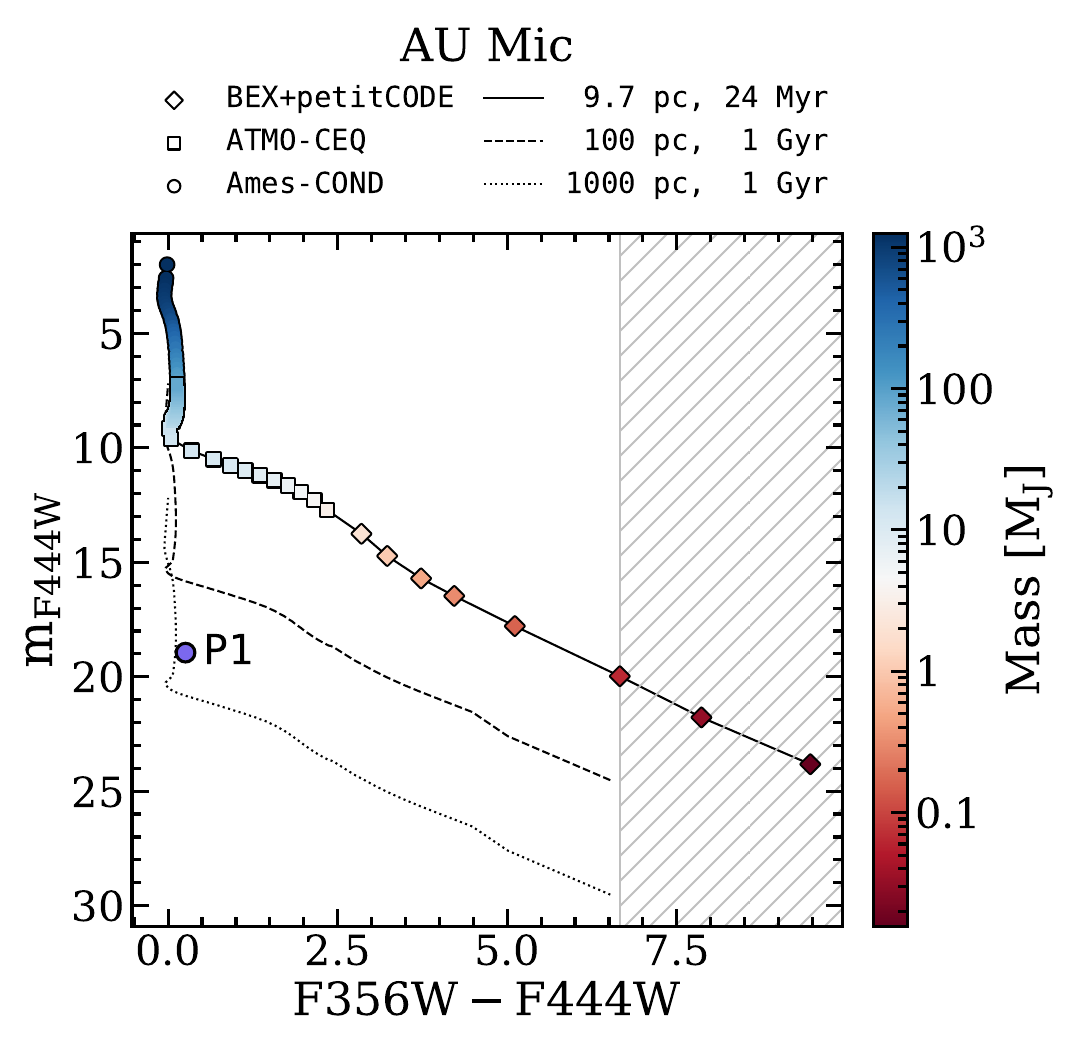}
    \includegraphics[width=0.33\linewidth]{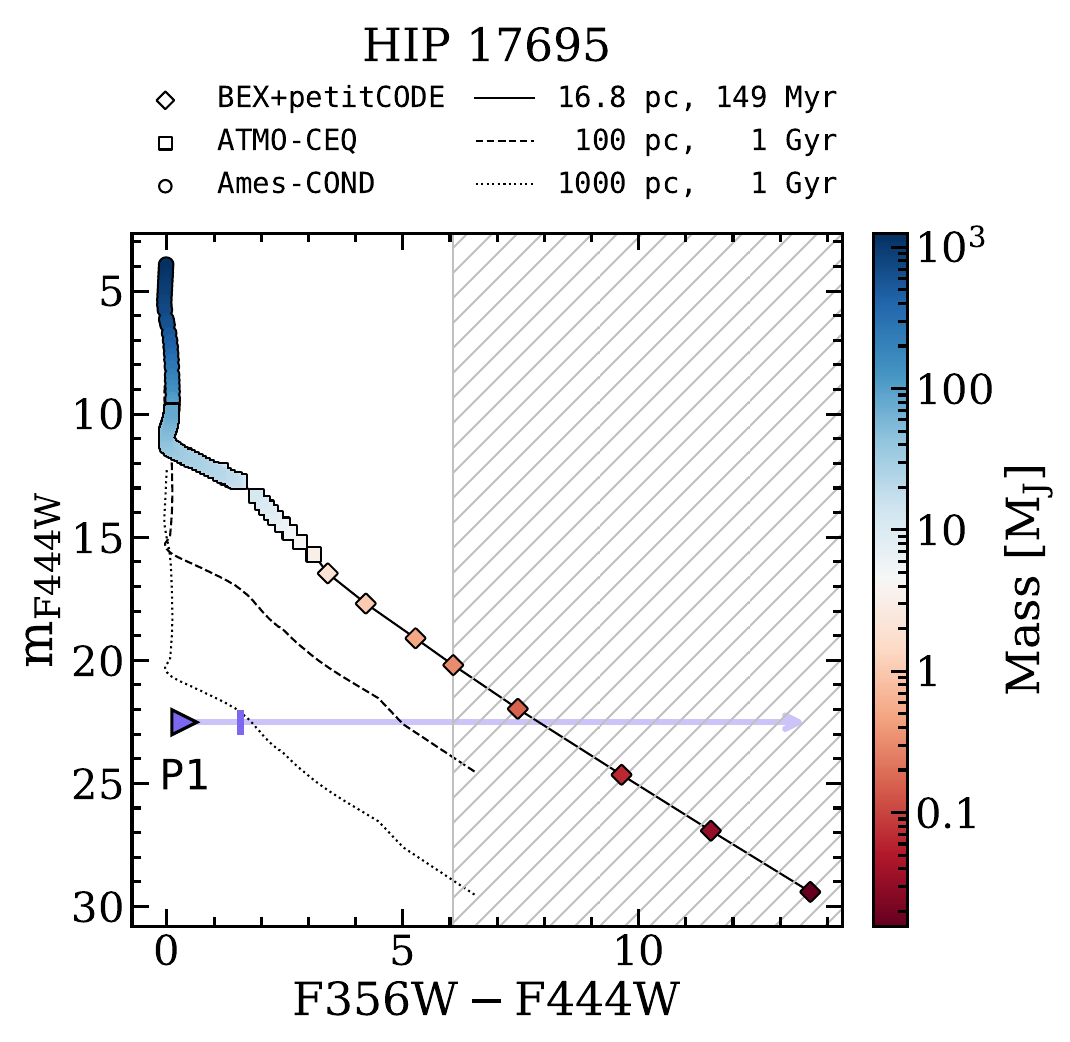}
    \includegraphics[width=0.33\linewidth]{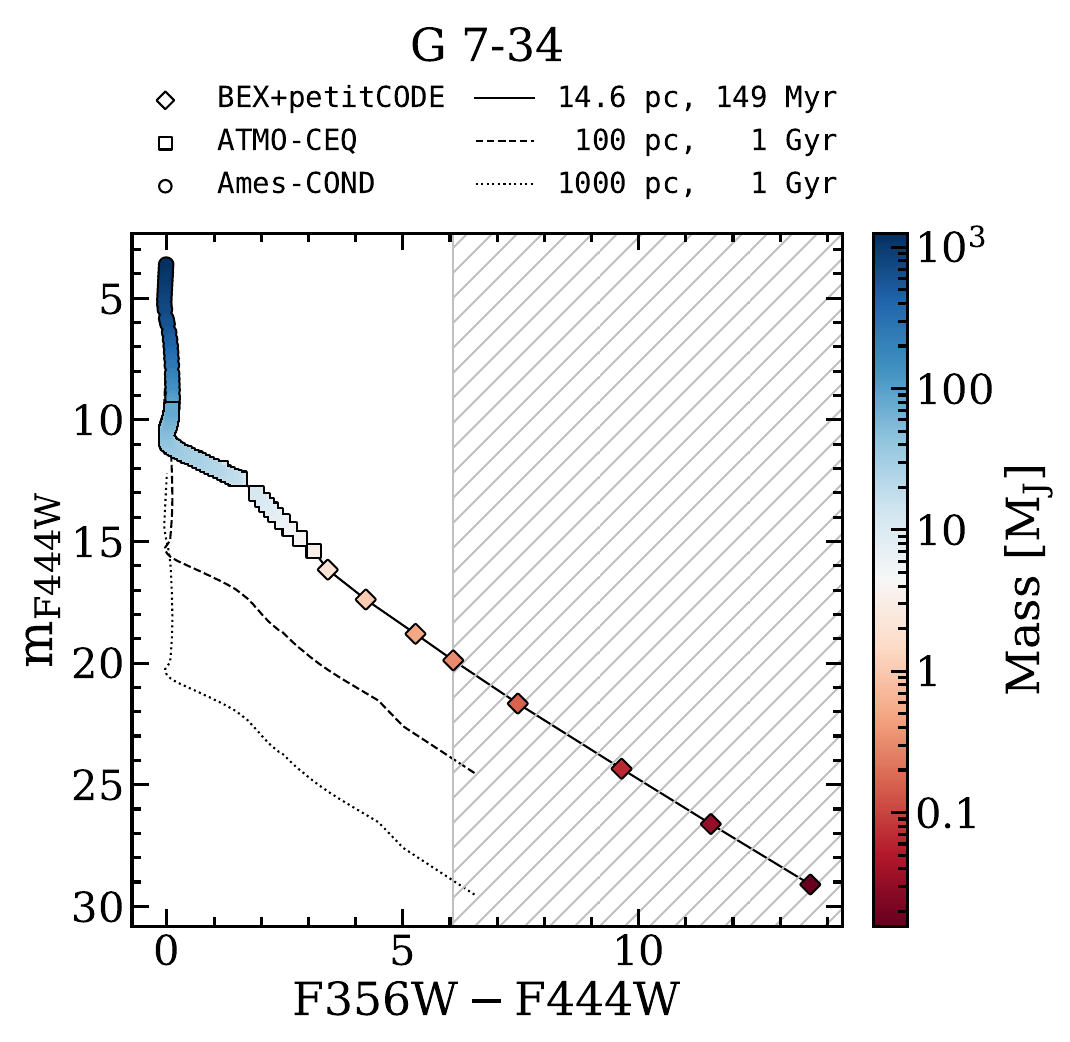}
    \includegraphics[width=0.33\linewidth]{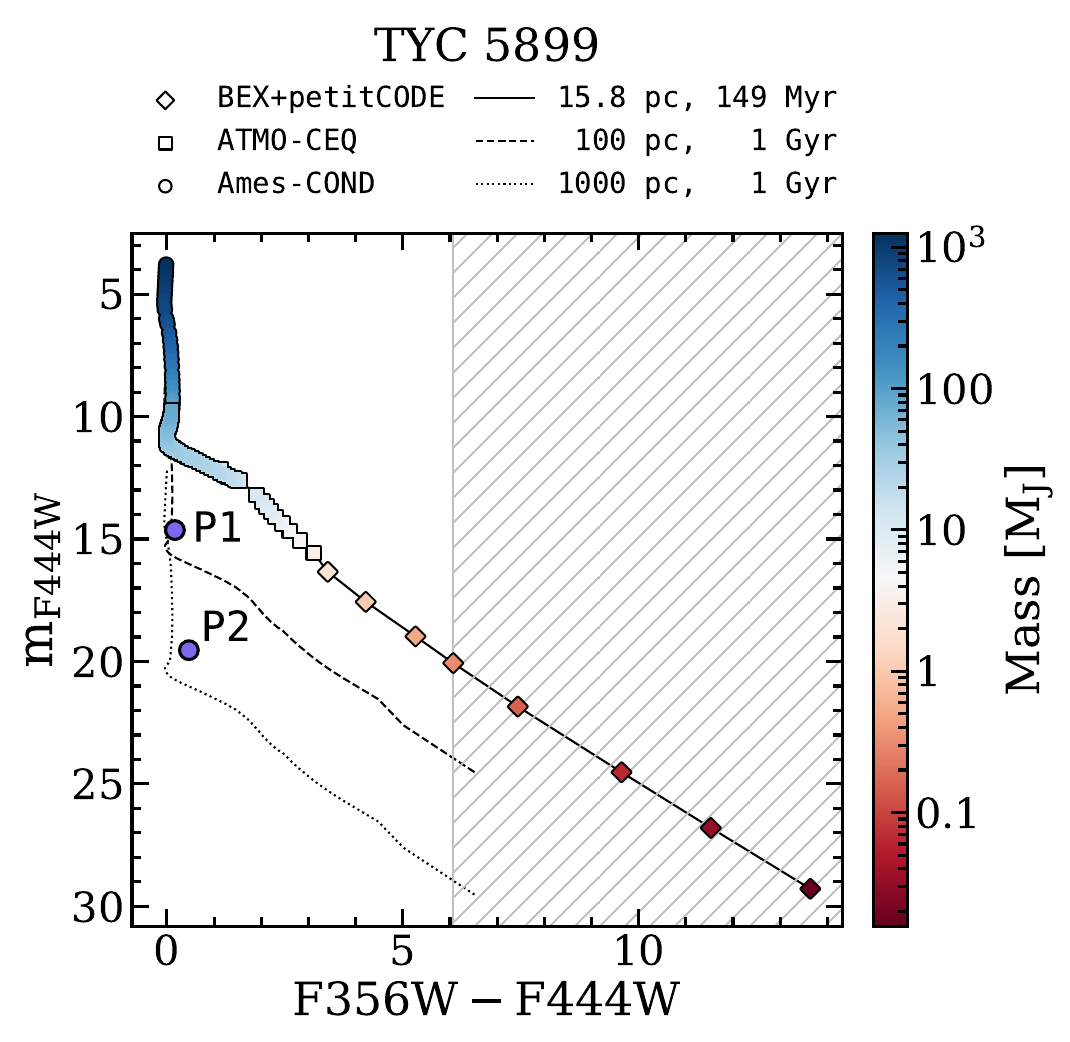}
    \includegraphics[width=0.33\linewidth]{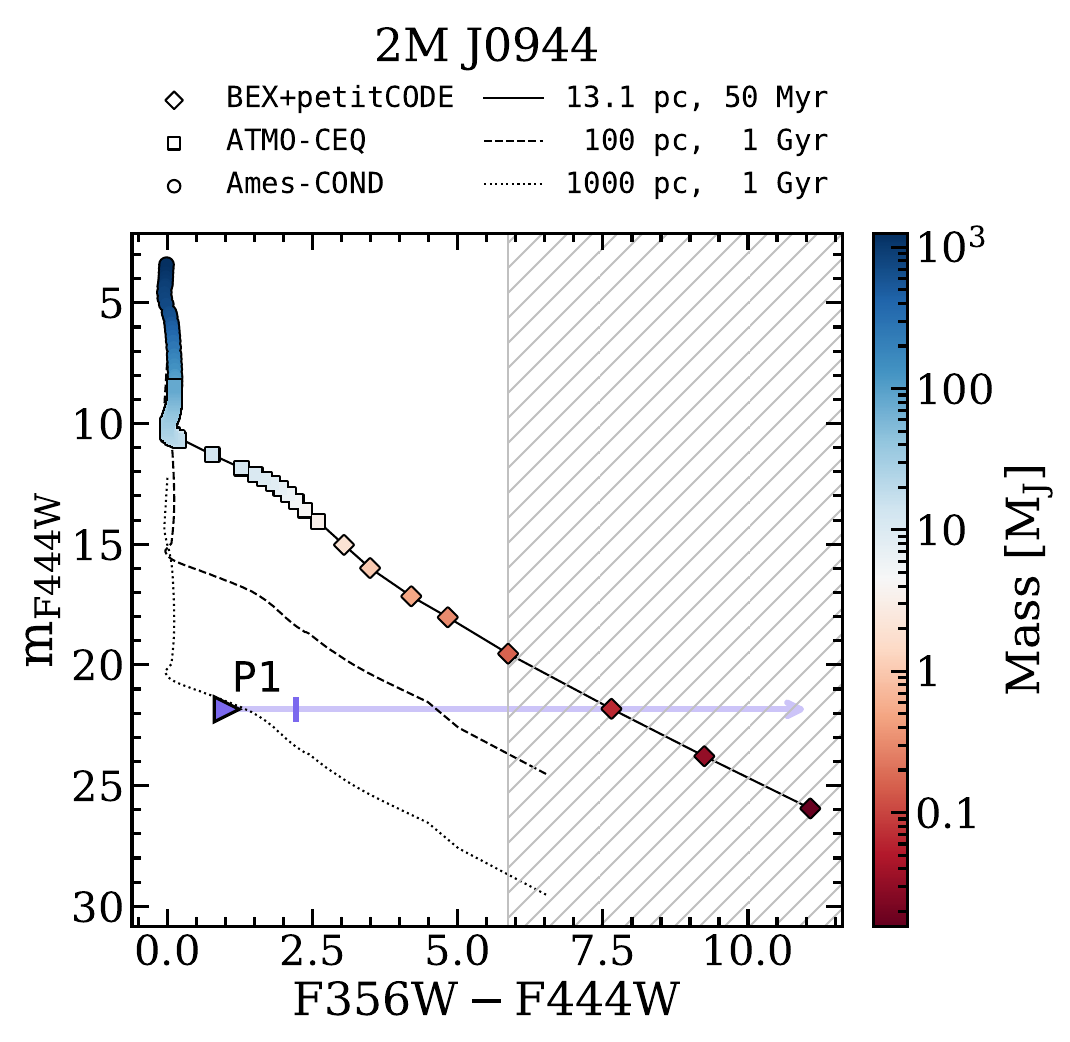}
    \includegraphics[width=0.33\linewidth]{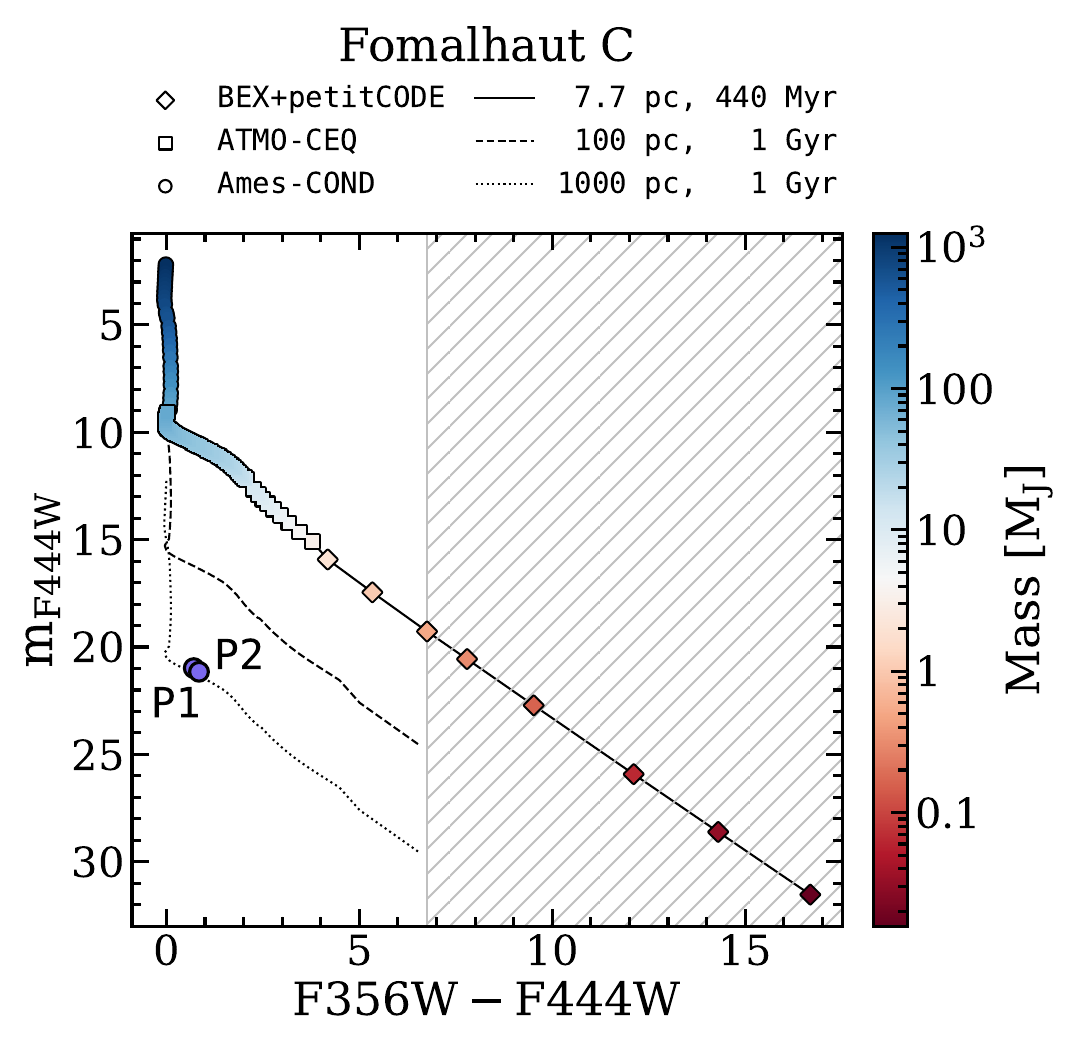}
    \includegraphics[width=0.33\linewidth]{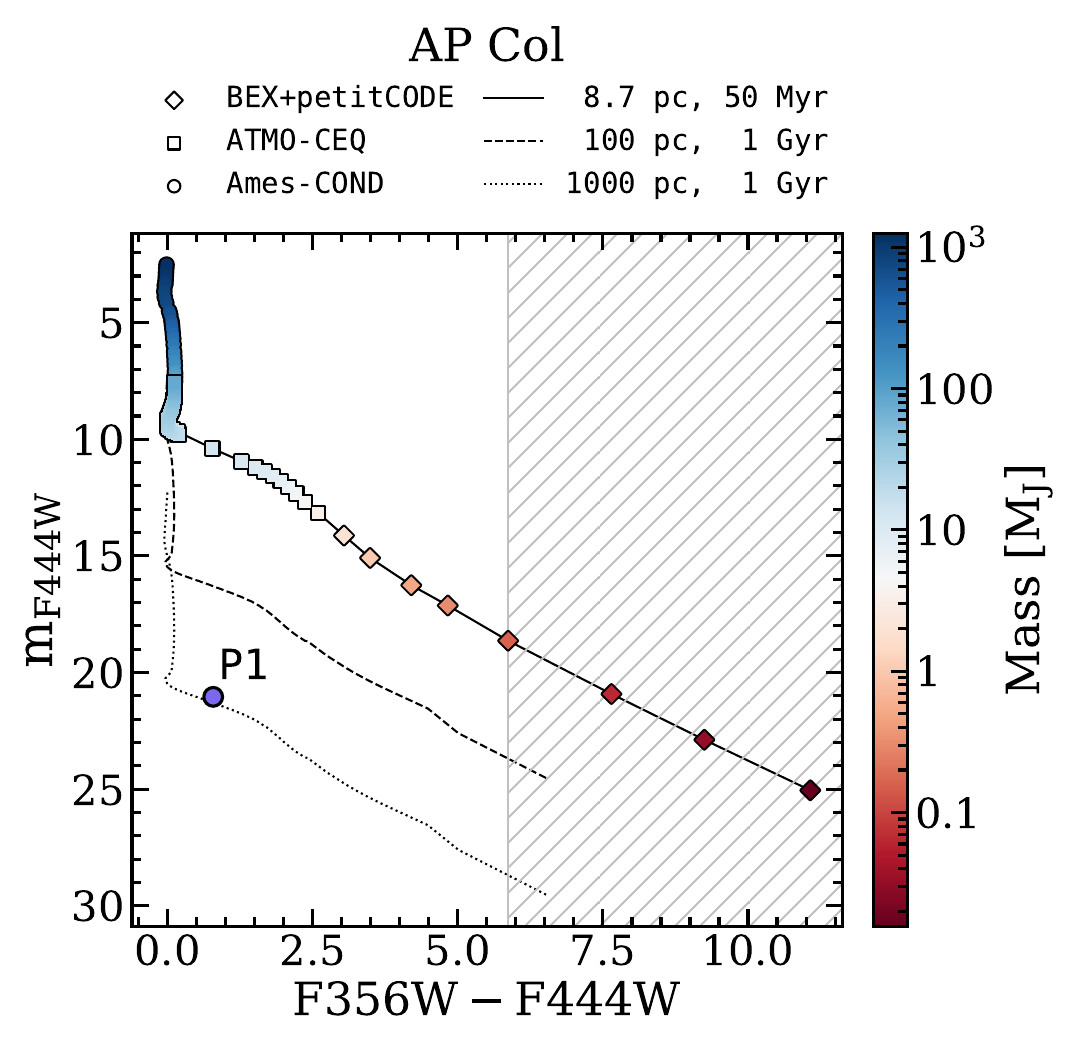}
    \includegraphics[width=0.33\linewidth]{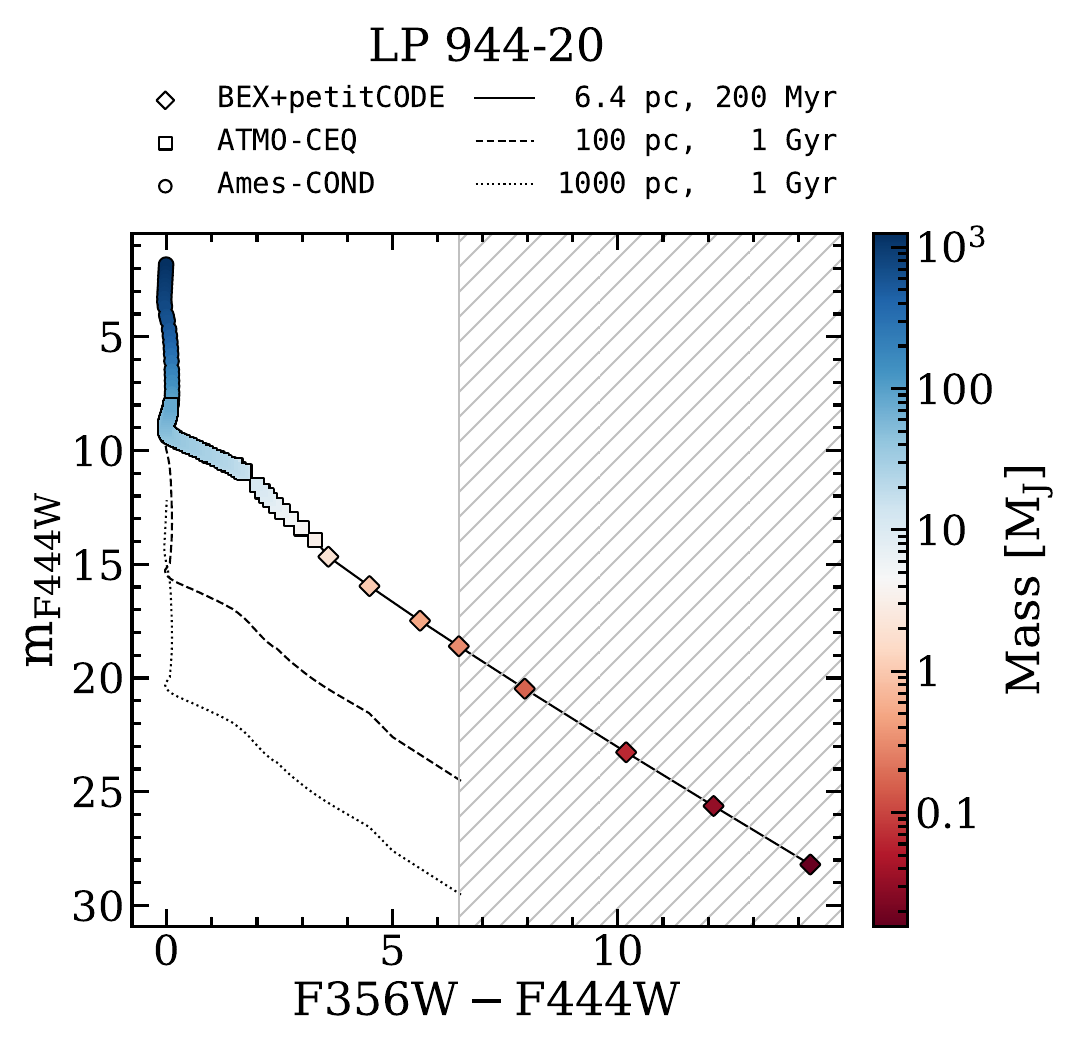}
    \includegraphics[width=0.33\linewidth]{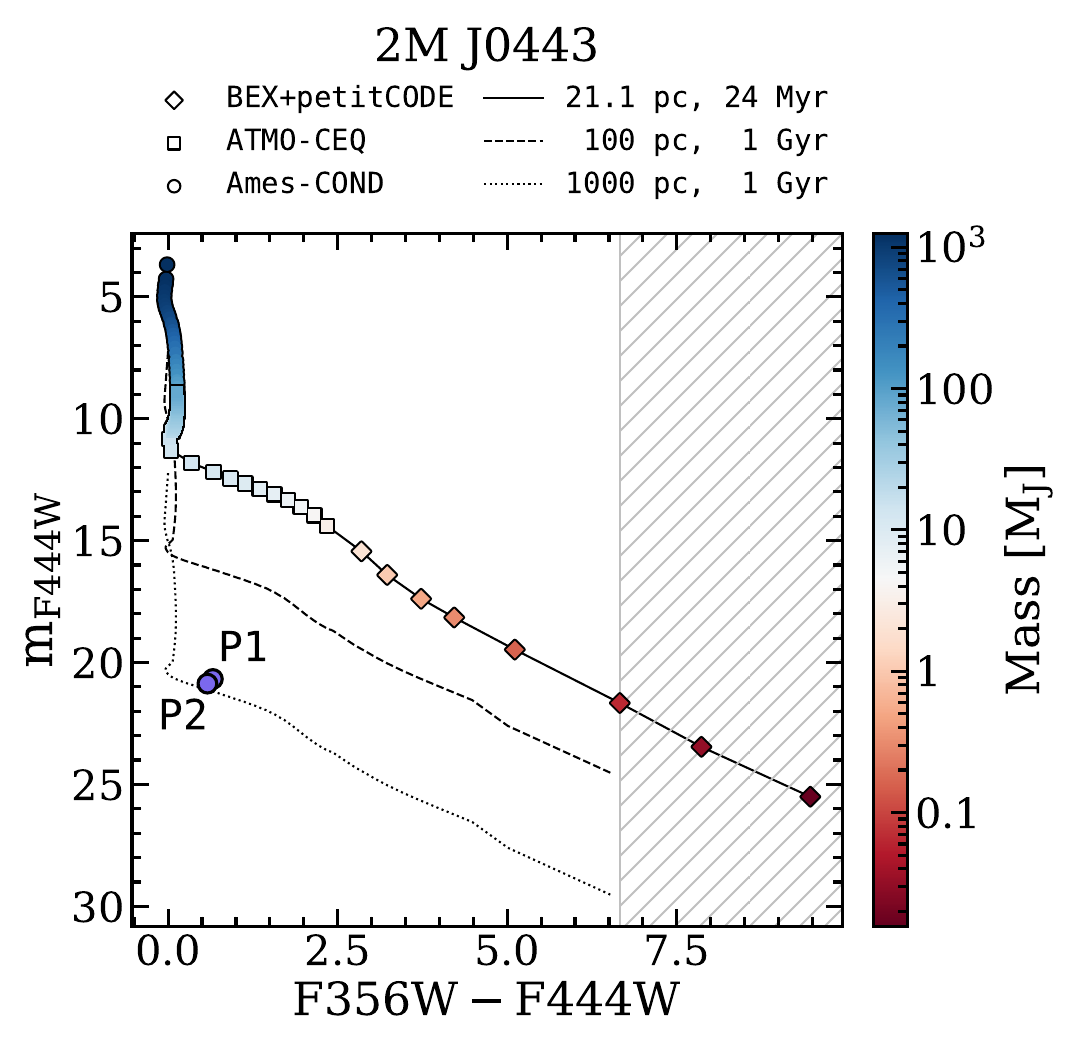}
    \caption{We compare the F444W photometry and F356W-F444W color of each detected point source (labeled purple icons) to the predicted photometry of stellar and substellar objects modeled at the age and distance of each target star (a.k.a. ``isochrone curve", shown in the solid line) in order to rule out background contaminants. The point sources detected in both filters (purple circles) are far removed from the isochrone and thus unlikely to be associated with the target system, however the two F356W dropout sources (purple triangle at the 3-$\sigma$ color limit, and vertical bar at the 1-$\sigma$ color limit) are consistent both with substellar field objects and in the limit of their colors, with a sub-Saturn-mass companion. The color of the marker on the isochrone curve corresponds to the object mass, and the predicted photometry is based on the \code{Ames-COND} (circle markers, for M $>$ 0.075 $\msun$), \code{ATMO-CEQ} (square markers, for  0.075 $\msun$ $>$ M $>$ 2 $\mjup$), and \code{BEX-petitCODE} (for M $\leq$ 2 $\mjup$) with the extrapolation to very low masses denoted by the hatched background. For comparison, we show isochrone curves for field age (1 Gyr) populations at 100 pc and 1000 pc in the dashed and dotted lines respectively.}
    \label{fig:isochrones} 
\end{figure*}

To determine if a given off-axis source is likely to be a companion associated with the target star, we compare the off-axis source photometry to an ``isochrone curve" based on the evolution model grids described in Section \ref{ssec:Image_Analysis}. This curve predicts the apparent magnitude in a given filter and the two-filter color for objects across a range of masses, assuming the same age and distance as the host star. Off-axis sources whose magnitude and two-filter color do not coincide with the target star's isochrone curve are most likely background objects and can be ruled out as potential companions. We show the detected point source position in relation to the system isochrones for each target star in the survey in Figure \ref{fig:isochrones}, along with 1 Gyr isochrones to represent possible field objects at distances of 100 and 1000 pc. Each point source which we detect in both filters (marked in the figure with a purple circle) is offset to significantly fainter magnitudes than the isochrone for its would-be host star given its colors, thus they can all be ruled out as more distant background sources. As mentioned previously, we obtained 3 total observations of TYC 5899 over a temporal baseline of 4 months. Given the high proper motion of this target, we were able to visually confirm that TYC 5899 P1 and P2, shown in Figure \ref{fig:isochrones} to be likely background objects, exhibited an apparent motion of approximately 0.5 pixels relative to TYC 5899 over the 4 months, confirming their status as background contaminants. 

Each F356W dropout source detected via SNRE is marked in the figure with a purple triangle at the position of the F444W photometry measurement and the $3\sigma$ lower limit on F356W-F444W color, as well as a vertical bar at the position of the $1\sigma$ lower limit, with an arrow extending to the right showing the range of colors possible. Comparing these results with the isochrones shown in Figure \ref{fig:isochrones}, we can see that the F444W magnitudes of the sources are fainter than the lowest planet masses available in the \cite{linder_evolutionary_2019} models for their respective target stars, but in the limit of their F356W-F444W colors they may be consistent with the color-magnitude space expected of a sub-Jupiter mass companion. If confirmed, these objects would be the lowest mass exoplanets ever imaged directly.

For both dropout sources, stellar background contaminants are ruled out by the positive F356W-F444W color limits. However, Figure \ref{fig:isochrones} shows that the available photometric constraints are also consistent with older isolated brown dwarfs (BDs) or planetary mass objects in the distance range of 100s of parsecs, and in some cases out to 1000 pc. Given the galactic latitude and longitude of each target star, we can quantify the probability of contamination by the Galactic disk population of isolated brown dwarfs (down to effective temperatures of $\sim$900 K) using the TRILEGAL model distribution \citep{girardi2005}, following the prescription used in \cite{deFurio2025}. We calculate the expected number of isolated brown dwarfs mimicking each dropout signal by counting the number of brown dwarfs in a cone extending in the direction of a given target star and which has a solid angle equal to that of an individual GTO 1184 observation ($\sim$400 square arcsec). We then filter for only the BDs with a predicted F444W magnitude less than the F444W magnitude upper limit of the dropout source as well as a predicted F356W magnitude greater than the flux limit of that dropout source. We present the expected number of contaminating background L dwarfs for each source, showing both the $1\sigma$ and $3\sigma$ F356W flux limit case in Table \ref{table:false_positives}. As our $1\sigma$ flux limits in F356W correspond to colors redder than those of the coolest brown dwarfs included in the TRILEGAL model (approximately spectral type T7), we effectively rule out brown dwarfs warmer than 900 K (a.k.a.~``warm BDs") and report 0 expected contaminating objects for that case. However in the $3\sigma$ flux limit case, we expect 0.021 contaminating warm BDs mimicking HIP 17695 P1 and a mere 0.001 contaminating warm BDs mimicking 2M J0944 P1.

\begin{table*}[t]
\centering
\begin{tabular}{ll|ccc}
\toprule
& & Expected Warm BDs & Expected Cool BDs & Expected Galaxies \\
Target & Source &  $3\sigma$ ($1\sigma$) &  $3\sigma$ ($1\sigma$) & $3\sigma$ ($1\sigma$) \\
\hline
HIP 17695 & P1 & 0.021 (0.0) & 0.003 ($0.002$) & 3.7 (0.012) \\
2M J0944 & P1 & 0.001 (0.0) & 0.001 ($4.3\times10^{-4}$) & 0.021 (0.004) \\
\toprule
\end{tabular}
\caption{ Here we show the expected number of contaminating background objects in a given observation that would mimic a dropout point source candidate, based on the $3\sigma$ F356W-F444W color limits (with results from the $1\sigma$ color limits in parentheses). The estimated warm ($\teff >$ 900 K) brown dwarf contamination is calculated from the TRILEGAL model of the galactic disk distribution of field substellar objects, the cool ($\teff <$ 900 K) brown dwarf contamination is based on an analysis of the 20 pc census in \cite{Kirkpatrick2024}, and the galaxy contamination is calculated from the observed population of galaxies in the JADES GOODS-S deep field.}
\label{table:false_positives}
\end{table*}

However, the GTO 1184 observations are sensitive to background objects significantly cooler than TRILEGAL's lower temperature limit of $\sim$900 K. In particular, brown dwarfs with T$_{\textrm{eff}}<900$ K (a.k.a~``cool BDs") are expected to have F356W-F444W colors greater than 1 magnitude and could still appear in our data as dropout signals. As these objects are not all included in the TRILEGAL model, we quantify the likelihood of their appearance by analyzing the volume-limited census of stellar and substellar objects within 20 parsecs compiled in \cite{Kirkpatrick2024}. We collect all the objects provided in the published database with effective temperatures below 900 K, ignoring those which are companions of more massive stars. We then bin by effective temperature, grouping objects between 900-750 K (67 objects), 750-500 K (99 objects), and 500-250 K (48 objects). For each of these categories, we calculate the mean number density per pc$^3$ within the observed volume, as well as the range of distances at which the objects would be detectable with a F444W magnitude less than a given dropout source but undetectable at F356W (i.e. having an apparent magnitude greater than the 3- or 1-$\sigma$ sensitivity limit in F356W at the separation of the dropout source). To do this, we use the provided WISE Band 1 and 2 magnitudes as a proxy for JWST F356W and F444W magnitudes respectively, given their similarity in wavelength coverage \citep{wright2010}. We then scale the number density of each category of BDs to the volume of space in which they would be observable as an object mimicking the dropout source according to our 3- and 1-$\sigma$ F356W detection limits. While this assumes the number density of cool BDs is uniform out to potentially 1.5 kpc (the furthest distance at which we could detect early T-dwarfs as dropout sources according to the 1$\sigma$ F356W limits), we expect the results displayed in Table \ref{table:false_positives} to be a conservative overestimate of the expected contaminating objects given that HIP 17695 and 2M J0944 are separated by $>100^\circ$ from the galactic center and by $>30^\circ$ from the plane of the galactic disk, where we would in reality expect the brown dwarf population density to decrease with distance. Even in this conservative case, we predict with $3\sigma$ confidence only 0.01 cool BDs to mimic HIP 17695, and only 0.001 cool BDs to mimic 2M J0944, given their intrinsic rarity and the narrow field of view of the NIRCam coronagraphic subarray.

\begin{figure}[t]
    \centering
    \includegraphics[width=\linewidth]{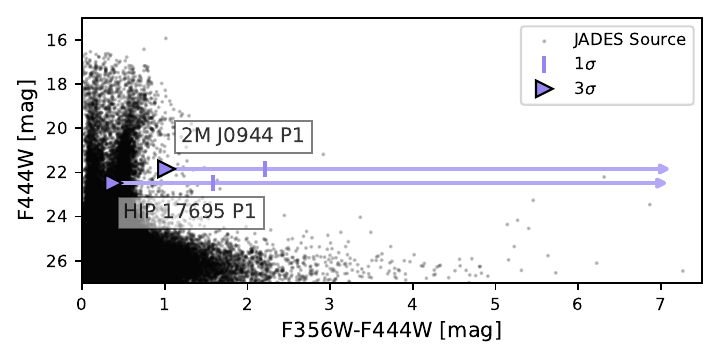}
    \caption{Color-magnitude diagram of the 2 dropout candidates (in purple) as compared to the sources in the JADES GOODS-S deep field catalog (in black), demonstrating the rarity of extragalactic sources with F444W magnitude $<24$ and F356W-F444W color $>1$. We show the $3\sigma$ and $1\sigma$ color limit of the dropout candidates with a triangle and verical bar, respectively. 2M J0944 P1 has the strongest color limit and a correspondingly low likelihood of being a background galaxy.} 
    \label{fig:jades}
\end{figure}

In addition, some galaxies may occupy the same magnitude and color space as a young, low-mass planet (as revealed by WISE and Spitzer observations), which is particularly problematic if they are also small and/or distant enough to appear point-like. Early-type galaxies have 3-5 µm colors comparable to stars \citep{wright_presence_2010}. However, low red-shift Quasars, obscured Active Galactic Nuclei (AGN), and Luminous Infrared Galaxies are redder, with quasars and AGN typically displaying colors ranging from $\sim$0.8 – 1.5 mags \citep{yan_characterizing_2013}. To determine the probability of contamination by such galaxies quantitatively, we interpret the early results of the JWST Advanced Deep Extragalactic Survey (JADES) \citep{robertson_identification_2023,curtis-lake_jades_2023,hainline2020}. JADES is in the process of observing 136 square arcminutes of sky to search for high redshift galaxies, with significantly deeper sensitivity in both F444W and F356W (down to $\sim$30th magnitude). The JADES NIRCam photometry catalog for the GOODS-S deep field \citep{eisenstein2023, rieke_jades_2023} represents a subsection of the full survey with the deepest limiting magnitude and covering an area of 27 square arcminutes. In Figure \ref{fig:jades} we show the F444W magnitude and F356W-F444W color of the JADES detections in black, as compared to the F444W magnitude and F356W-F444W color limits of the GTO 1184 dropout sources in purple (with a triangle at the $3\sigma$ limit and a vertical bar at the $1\sigma$ limit). We count the number of sources in the catalog that have a F444W magnitude less than or equal to to the upper limit of a given GTO 1184 dropout source, as well as a F356W magnitude greater than the F356W magnitude limit for the GTO 1184 source, as in the brown dwarf false positive calculation. We then scale by the sky coverage of the GOODS-S deep field compared to the GTO 1184 survey field-of-view (0.11 square arcminutes per target), and determine the expected number of JADES sources that would mimic the dropout source in its field of view. As shown in the rightmost column of Table \ref{table:false_positives}, the expected number of contaminants for source 2M J0944 P1 is a scant 0.021, and we gain confidence in ruling out background galaxy contaminants as the source of this signal. However, HIP 17695 P1 has a $3\sigma$ color limit consistent with more frequently occurring galaxies, therefore background contamination is more likely. We note that while the results for HIP 17695 imply that we should see several dropout galaxies in that FOV (the expected number is 3.667), the vast majority of planet-mimicking galaxies in the JADES data lie right near the $3\sigma$ detection limit in F444W for GTO 1184. Thus our detection of only a few dropout candidates with an SNRE of 4 or more is consistent with the expectation.

\section{Discussion} \label{sec:Discussion}

This work represents the first probe of the sub-Saturn-mass exoplanet population via direct imaging, thus we have a responsibility to explore even marginal detections which are potentially consistent with low-mass planetary companions. The first dropout candidate, HIP 17695 P1, was initially detected with an SNRE of 4.84 in F444W. This was calculated by comparing the flux within the FWHM of a given point to the noise level averaged on an annulus surrounding the star at that separation. However after fitting a PSF model to the source and comparing the resulting flux to the local noise level, we found that that the SNR is only 1.64. This, in combination with the higher likelihood of contamination by extragalactic sources as shown in Table \ref{table:false_positives}, makes it less compelling and leads us to demote it from consideration as an exoplanet candidate.

The remaining dropout source, 2M J0944 P1, with a marginally significant PSF fit SNR of 3.91 in F444W per Table \ref{table:dropout_photom} and an expected number of contaminating background sources of 0.023 objects with $3\sigma$ confidence, remains an intriguing candidate. The simplest course of action to confirm or reject the exoplanet nature of this source is to re-observe the system, re-detect the source at higher SNR, and verify common proper motion with the host star. Each of our target stars has a proper motion on the order of 100s of mas yr$^{-1}$ or more, and common motion would be straight forward to detect after only one year given the astrometric accuracy for faint sources of $\lesssim$20 mas with NIRCam as demonstrated in this work.

Follow-up observations of the 2M J0944 system were recently executed in JWST Cycle 2 Guest Observer Program 3840 and the results of those analyses will be published in a future paper (Bogat et al., in prep). We note that the original candidate for this followup program was a $5\sigma$ source in the vicinity of G 7-34, however it was revealed to be an artifact due to imperfect cosmic-ray correction with re-reduction of the full dataset using updated versions of the JWST pipeline and \code{spaceKLIP} (see Section \ref{ssec:Data_Reduction}). We were able to swap G 7-34 with 2M J0944 through a target change request, as 2M J0944 hosted our second-best dropout point source candidate at that time.

We also note that no objects consistent with Jupiter-mass, super-Jupiter-mass, or BD companions were detected in the survey. This is loosely in keeping with previous results showing low occurrence rates for giant planets orbiting M dwarfs \citep{ribas_2023}, and we provide quantitative context for these results in the following subsection.

\begin{table*}[t]

\centering

\begin{tabular}{lll|ccc}
\toprule
\textbf{Category} & \textbf{Mass} & \textbf{SMA} & \multicolumn{3}{c}{\textbf{Frequency Upper Limit}} \\
 & \textbf{Range} & \textbf{Range} & ($1\sigma$) &  ($2\sigma$) & ($3\sigma$) \\
\hline
\textbf{Sub-Jupiters} (log-norm)  & 0.1-1 $\mjup$ &  10-200 AU & $<0.10$ & $<0.13$ & $<0.16$ \\
\textbf{Sub-Jupiters} (log-flat) & 0.1-1 $\mjup$ &  10-200 AU  & $<0.10$ & $<0.14$ & $<0.17$ \\

\textbf{Super-Jupiters} & 1 $\mjup$ - 0.1 M$_{*}$ &  10-200 AU & $<0.12$ & $<0.15$ & $<0.21$ \\
\textbf{Brown Dwarfs} & 3-75 $\mjup$ &  10-200 AU & $<0.16$ & $<0.18$ & $<0.23$ \\
\toprule
\end{tabular}

\caption{Occurrence constraints (objects per star) for substellar companions in different mass categories with semimajor axes between 10-200 AU that would result in the expectation of 0 detections in the survey. The sub-Jupiter results are shown for both the case where the underlying separation distribution is a log-normal distribution centered at 4 AU (denoted ``log-norm'' in the table), and the case where it is represented by the same log-normal distribution within 10 AU but transitions to log-flat beyond 10 AU (denoted ``log-flat'' in the table). The estimates are generated by simulating the substellar companion population described by Meyer et al. 2025 (in prep) and comparing with the 68\%, 95\%, and 99.7\% detection probability contours (see Fig. \ref{fig:combined_detection}) for the 1-, 2-, and 3-$\sigma$ confidence estimates respectively. The super-Jupiter and brown dwarf constraints are loose but consistent with previous direct-imaging surveys. The results for sub-Jupiters are the first occurrence constraints produced in this mass-separation regime, and are not significantly sensitive to the underlying separation models given the small sample size.}
\label{table:occurrence}
\end{table*}

\subsection{Implications for Population Statistics}
\label{ssec:occurrence}

The existing observational constraints for giant M dwarf planets with masses 1-10 M$_{\rm Jup}$ are consistent with a single underlying planet population, however several possible functional forms may accurately describe the planet surface-density distribution. \citet{Clanton2016a} adopted a power-law with a sharp outer cutoff radius around 10 AU to match imaging upper limits at the widest separations. \citet{Meyer2018} included newer imaging constraints and found that a log-normal distribution with a peak around 3 AU fit the extant observations. However, due to the limited mass sensitivity of the previously available high-contrast imaging observations, the population of sub-Jupiter-mass planets at wide separation lacks constraint. 

To explore the population in this newly accessible mass regime, we perform yield simulations based on a recent parametric demographics model for M dwarf companions (Meyer et al. 2025, in prep) and our sensitivity constraints from the 5$\sigma$ sensitivity curves for each target (see Appendix \ref{a:contrast}). We simulate the yield for brown dwarf binary companions, as well as for planets in two different planetary mass regimes, dividing the planet population into sub-Jupiters and super-Jupiters with the transition defined at 1 $\mjup$. Each population is represented by a log-normal orbital distribution (peaking at $\sim$4 AU for sub- and super-Jupiters, and $\sim$30 AU for brown dwarfs) and a power-law mass ratio distribution (index of $\sim$-1.43 for planets and $\sim$0.3 for brown dwarfs). The super-Jupiter model is constrained by empirical evidence from previous exoplanet demographics surveys, including the HARPS radial velocity survey \citep{grandjean2023}, SPHERE SHINE survey \citep{vigan2021}, and GPIES \citep{nielsen_gemini_2019}. We set the lower mass limit for planets at 0.03 $\mjup$ (or 10 M$_{\oplus}$), as this is below the median detection threshold for the survey. The upper mass limit for planets is defined as 10\% of the stellar mass, instead of the deuterium-burning limit of 13 $\mjup$, as M dwarf disks are not expected to be massive enough to create 13-Jupiter-mass objects via a planet-like formation pathway. For companion generation, we draw $10^5$ random sets of orbital parameters using the following priors: uniform priors for the longitude of the ascending node and the longitude of periastron, cosine priors for inclination, and Gaussian priors for eccentricity ($\mu = 0$, $\sigma = 0.3$; \citep{Hogg2010}), and consider a given planet ``detected" if it appears above the 5-$\sigma$ contrast curve for its host star. We calculate the expected number of planets to be detected around each star individually, assuming a total frequency (objects per star between 0 and 200 AU) of 0.02 for BDs, 0.09 for super-Jupiters, and 0.18 for sub-Jupiters, then sum the results to achieve an expected yield for the entire program. This results in an estimated yield for GTO 1184 of 0.45 $\pm$ 0.13 sub-Jupiters, 0.1 $\pm$ 0.012 super-Jupiters, and 0.085 $\pm$ 0.05 brown dwarfs. 

We also explore an alternative sub-Jupiter model, which follows the same mass ratio distribution as the gas giants but features a modified separation model. This model assumes the same log-normal distribution of semimajor axis as that of the super-Jupiters within 10 AU \citet[][c.f.]{fulton_2021}, but transitions to a log-flat distribution beyond 10 AU. The extension is motivated by the presence of the Solar System outer planets and by recent mm-wave ALMA disk gap observations \citep[e.g.][]{Garufi2024}. Under this assumption, the expected program yield of sub-Jupiters increased to 1.36 $\pm$ 0.18, effectively tripling the previous expectation.

In the super-Jupiter and BD regimes, our predictions are consistent both with the lack of detections in GTO 1184 and with the low yields from previous, less-sensitive M-dwarf observations from the ground \citep[e.g.][]{bowler2015}. For sub-Jupiters, a log-flat distribution of objects beyond 10 AU leads to a significant boost in the yield prediction at wide separation compared to the higher mass companions, and our identification of the marginal sub-Jupiter candidate 2M J0944 P1 is intriguing in this context. However, the small sample size of the survey, in combination with parametric degeneracies in the occurrence model, prevents us from definitively determining whether the log-normal or log-flat distribution of sub-Jupiters beyond 10 AU more accurately represents the underlying population via these observations alone.

Assuming conservatively that 2M J0944 P1 is not a planet, we can place an upper limit on the frequency of planetary companions in the mass and separation range covered by GTO 1184. We simulate 10,000 surveys with varying amplitudes of the planet and BD occurrence distribution, using the log-flat distribution of sub-Jupiters, and compare the simulated populations with the detection probability curves to determine the maximum possible planet and brown dwarf frequency (a.k.a. the intrinsic number of objects per star) that still results in an expectation of 0 detections in our observations. We adopt the 68\%, 95\%, and 99.7\% detection probability contours respectively (shown in Figure \ref{fig:combined_detection}) and find that the frequency of planets with masses between 0.3-1.0~$\mjup$ and semimajor axes between 10-200 AU (reflecting our sensitivity limits) is $< 0.10$ with $1\sigma$ confidence, $< 0.14$ with $2\sigma$ confidence, and $< 0.17$ with $3\sigma$ confidence. When we repeat the analysis assuming the log-normal distribution of sub-Jupiters beyond 10 AU, we find very similar limits of $< 0.10$ with $1\sigma$ confidence, $< 0.13$ with $2\sigma$ confidence, and $< 0.16$ with $3\sigma$ confidence, showing that the occurrence constraints are not significantly model-dependent. These results, as well as the occurrence constraints for the super-Jupiter and BD populations, are summarized in Table \ref{table:occurrence}. Given the small sample size of our survey, the constraints are necessarily loose. However, continued imaging of nearby, young M dwarf systems will build up the statistical power of our early observations to provide a clearer picture of wide-orbit giant planet occurrence around M stars. 


\subsection{Lessons Learned for JWST NIRCam Coronagraphic Imaging}

In GTO 1184, we achieved a flux contrast roughly similar to the results of the Direct Imaging ERS observations \citep[F444W $5\sigma$ contrast of $1.5 \times 10^{-5}$ at 1" in this work as compared to $1.0 \times 10^{-5}$ at 1" in][]{carter_jwst_2022}, but our choice of cooler target stars results in deeper companion mass sensitivities for individual targets by up to a factor of 10. Notably, we achieve this without the use of dedicated reference stars, thereby maximizing the survey efficiency and the total observation time used for science targets. 

We also opted not to use the small-grid dither approach \citep{Lajoie2016}, which is commonly employed when using dedicated PSF reference stars to increase the PSF sample variety. However, dithering on a science target is not recommended as it would result in the majority of the exposures having a sub-optimal star-coronagraph alignment. As noted in Figure \ref{fig:SubtractionContrast}, we found that RDI without dithering can outperform the sensitivity of the pure ADI reduction within the IWA even with only a single reference star used, but that this depends on the star-mask alignment of the reference observation being closer to that of a given science observation than the two science roll images are to each other. In the background-limited regime, we find that the inclusion of additional science targets as PSF references further improves the sensitivity. We note that the masking applied to prepare the science observations for use as PSF references (described in Section \ref{sssec:psf_library}) was critical to allow the maximum number of references to be useful without generating subtraction artifacts. Further contrast sensitivity improvements may be possible by incorporating PSFs into the reference library from other NIRCam coronagraphic surveys, including JWST GOs 4050 and 5835 (PI Carter) and GO 6005 (PI Biller), however these data were not publicly available at the time of our analysis.

We note that we chose to have larger slews between adjacent observations (occasionally $\gtrsim 50^\circ$) in order to group target observation sequences by similar spectral type. Due to the exceptional thermal stability of the observatory, these slew distances did not have a significant effect on the WFE drift between adjacent observations \citep[see][]{McElwain2023}, and therefore did not adversely affect the PSF subtraction quality. 

\subsection{Caveats \& Limitations}
\label{ssec:caveats}

There are several as-yet undiscussed factors which add uncertainty to our results. The first is that we rely on a single, cloud-free model grid \citep[\code{BEX-petitCODE},][]{linder_evolutionary_2019} for predicting the photometry of planets below 2 $\mjup$, which may not accurately reflect their physical conditions. Comparisons to models which include clouds \citep[e.g. \code{BT-SETTL},][]{allard_2013} and/or disequilibrium chemistry \citep[e.g. \code{ATMO 2020-NEQ},][]{phillips_new_2020} would provide a more complete picture of the possible planet parameters compatible with a given detected source. However, no atmospheric models currently exist which include self-consistent prescriptions for both clouds and disequilibrium chemistry simultaneously, especially near or below effective temperatures of 150 K. 

Furthermore, the \code{BEX-petitCODE} model grid has a low temperature limit of 150 K, while we demonstrate that JWST NIRCam coronagraphy is sensitive to planets significantly fainter, thus lower mass and presumably colder than can currently be modeled. We have performed a basic extrapolation (linear in log space) to be able to visualize the most likely accessible planet masses in our survey (see Figure \ref{fig:combined_detection}). Around the faintest and youngest target in the GTO 1184 sample (2MJ 0443), we achieve model predicted sensitivity to planets comparable in mass and separation to Neptune in our Solar System. These results reveal that extensions to exoplanet evolution and atmosphere model grids that include the relevant physics for ice-giant like planets at temperatures $<$150 K are needed to fully interpret NIRCam's sensitivity. 

We also note that our differentiation between point-like and extended sources was performed by inspecting PSF fit residuals by eye, rather than by an explicit quantitative metric. The ability to quantify the ``extendedness" of a given source is being implemented into \code{spaceKLIP} at the time of writing, and will be useful to better rule out extended sources in the future. 

We reiterate that we reduced our SNRE threshold to yield the detection of several F356W dropout sources. Assuming gaussian errors, one might expect 1 in 15800 pixels to produce a noise fluctuation above a 4$\sigma$ level. The field of view of each observation is roughly 100,000 pixels, so it is possible that a few noise speckles may masquerade as astrophysical sources. As we have cross-checked each dropout source to ensure the signal remains above SNRE 4 across multiple reduction strategies, we believe this chance is reduced (indeed several initial dropout candidates were ruled out this way) however nonzero. Given our 4$\sigma$ F444W SNRE cutoff for these sources, lower limits on the F356W-F444W colors, and the lack of diagnostic follow-up, we do not claim a definitive detection of a sub-Saturn-mass exoplanet. However, these results indicate that deeper observations of 2M J0944 would provide a more significant detection as well as better color limits, and they have been pursued in the aforementioned GO 3840 follow-up observations. 

As alluded to in Section \ref{ssec:occurrence}, both the small sample size of the GTO 1184 survey and the presence of parametric degeneracies in the occurrence model prevent us from definitively determining whether the separation distribution of sub-Jupiters at wide orbits (beyond 10 AU) follows a log-normal or log-flat distribution. The effect of the small sample size is self-evident, and we note explicitly that even if the sub-Jupiter candidate 2M J0944 were confirmed, we would still need more high-contrast imaging observations to meaningfully constrain the occurrence rate of sub-Jupiters beyond 10 AU. Regarding degeneracies in the model, currently both the peak amplitude of sub-Jupiter frequency (at 4 AU in either case) and the shape of the separation distribution's wide-orbit tail are ill-constrained by observations. The frequency for the log-normal distribution is predicted to be 0.18 objects per star across separations from 1-100 AU. Assuming the same frequency within 10 AU but transitioning to a log-flat distribution beyond 10 AU results in approximately triple the expected sub-Jupiters at wide orbits. Thus, if we measure a frequency of wide-orbit sub-Jupiters to be significantly higher than expected from the log-normal distribution case, we might deduce that the underlying population must indeed have some extended tail. However it could also be explained by the sub-Jupiter population at all separations being elevated, but still having a log-normal separation distribution. \cite{Poleski2021} summarizes $\sim$20 years of microlensing data from the Optical Gravitational Lensing Experiment (OGLE), finding that each microlensing star hosts $1.4^{+0.9}_{-0.6}$ ice giants (with semimajor axis 5-15 AU and planet-star mass ratios ranging from $10^{-4}$ to $3.3\times10^{-2}$) on average. To confidently differentiate between the two wide-orbit demographic models, we will need stronger constraints on the sub-Jupiter population in the inner (1-10 AU) separation region around M dwarfs, which are expected to be provided by exoplanet discoveries from the Roman Galactic Bulge Time Domain Survey \citep{Penny2019,Wilson2023} and \textit{Gaia} astrometry in the next several years \citep{Sozzetti2014}.

\section{Summary \& Conclusions}
\label{sec:Conclusions}

We present results from JWST Cycle 1 GTO Program 1184, designed to survey a carefully selected sample of very nearby, young, M dwarfs with JWST NIRCam Coronagraphy at 3 - 5 $\mu$m wavelengths to search for sub-Jupiter mass planets on wide orbits. The key approaches and results of the survey are as follows:

\begin{itemize}

  \item {We chose very nearby, young, low-luminosity M dwarfs as targets to access planets and disks at Solar System-like separations and achieve very deep planet mass sensitivity in contrast limited observations;}

  \item {We used the standard high-contrast imaging post-processing approach of KLIP RDI and the open source software package \code{spaceKLIP} to construct optimized PSF references using the science target observations (a self-referenced survey). This approach allowed us to perform an efficient survey while reaching contrasts consistent with other JWST NIRCam observations that observed dedicated PSF references;}

 \item {Our observations and post-processing approach result in reduced images with typical model derived sensitivities to Saturn-mass planets at Saturn like projected separations (9.5 AU). For some targets, the sensitivities push down to approximately the mass of Neptune at $\gtrsim$10 AU. We are able to rule out companions with masses $\gtrsim 0.3~\mjup$ in the 10s of AU region around each of the 9 target stars;}

 \item {While our model derived mass sensitivities are the deepest yet achieved via direct imaging observations, we note that the GTO 1184 observations are sensitive to masses below the lower limits of the available planet evolution model grids at the young ages of the targets. The models need to be extended to lower masses (cooler temperatures) to fully explore the available parameter space without extrapolation;}

 \item {We identify a marginal source, 2M J0944 P1, that has a 3-5 $\mu$m photometric color limit consistent with the expectation for a $\sim$Saturn-mass planet. For this source, we estimate low probabilities of background contamination by stars, field brown dwarfs, or galaxies. In a future publication, we will present a proper-motion analysis of recently collected follow-up NIRCam coronagraphic imaging data to either confirm or refute the association of 2M J0944 P1 with 2M J0944.} 

 \item{We also place the survey results in context with planet population models in three different mass regimes. In particular, we place the first occurrence rate constraints on wide-orbit, sub-Jupiter (semimajor axes 10-200 AU, masses 0.3-1~$\mjup$) exoplanets around M dwarfs to be $<0.10$ and $<0.16$ objects per star with 1- and 3-$\sigma$ confidence respectively, assuming the Meyer 2025 (in prep) population model;}

  \item {Finally, GTO 1184 allowed for the first detection of the well studied AU Mic debris disk at 3-5 $\mu$m \citep{Lawson2023} and the first reflected light detection of the faint disk associated with Fomalhaut C \citep{Lawson2024}.}

\end{itemize}

The execution and analysis of GTO 1184 survey data has further demonstrated the unprecedented sensitivity of JWST NIRCam coronagraphy and its potential to advance exoplanet science. The search for wide-separation planets in young M dwarf systems provides insight into the broad population of their exoplanets from the inner regions to outer reaches of these systems. The deep, wide-orbit constraints provided by GTO 1184 also provide context for the orbital and thermal evolution of the inner and potentially habitable planets of these systems. With several larger JWST surveys currently planned to search for planetary companions of young, low-mass stars, e.g. GO 4050/5835 (PI Carter), GO 6005 (PI Biller), GO 6122 (PI Bowens-Ruben), GO 8826 (PI Lawson), we have entered a new regime of direct sub-Jupiter imaging with JWST.


\appendix

\pagebreak
\section{Flux Maps}\label{a:flux_maps}

In our primary analysis, we use the SNRE maps as the key diagnostic tool, however the measurement of per-pixel noise and individual source photometry is done in units of flux (MJy/steradian). Here we show the images of each target in the F444W filter for reference, and the SNRE maps of each target can be found in Section \ref{sec:Results}.

\begin{figure*}[b]
    \includegraphics[width=\linewidth]{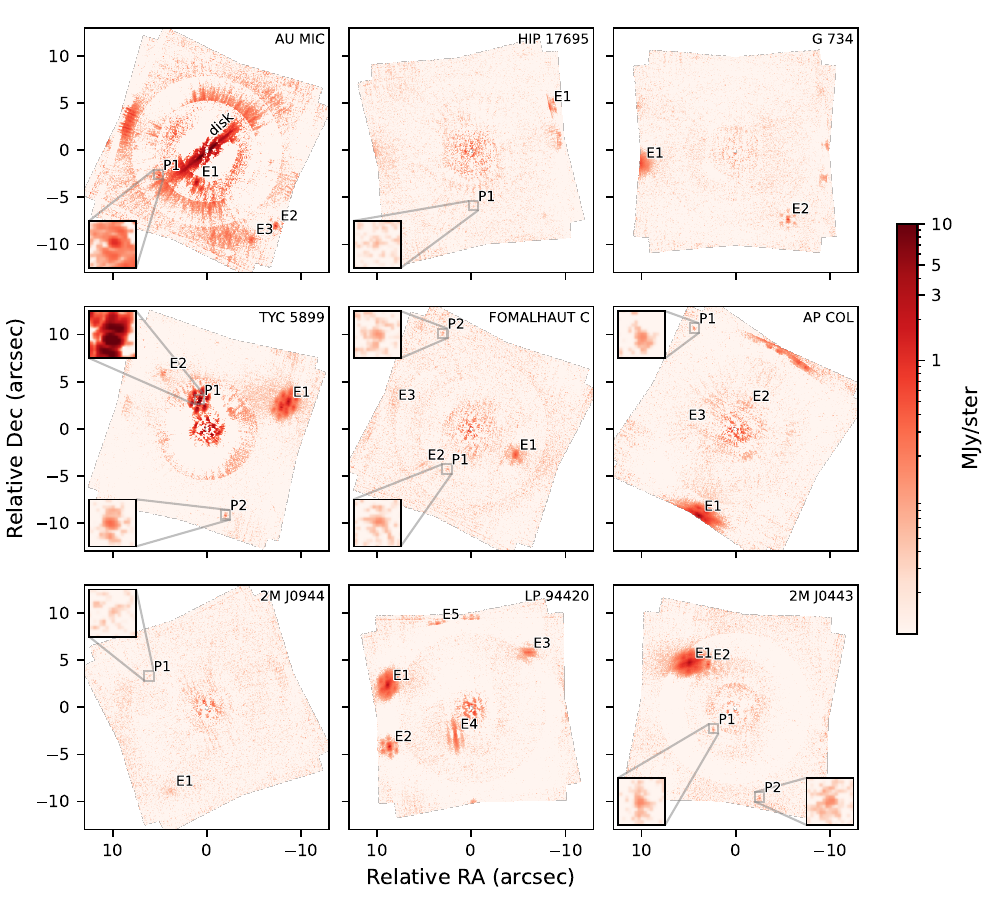}
    \caption{Flux maps of all targets in GTO 1184, in the F444W filter. All observations were reduced with RDI using the full library of reference PSFs. Point-like and extended sources are labeled in the images (with names beginning with ``P" and  ``E" respectively), and the point-like sources are magnified in the image insets. Extended background sources are abundant in the survey, and the edge-on disk of AU Mic is clearly detected.}
    \label{fig:flux_mosaic_f444w}
\end{figure*}

\pagebreak
Here we show the images of each target in the F356W filter (in flux units) for reference. The SNRE maps of each target can be found in Section \ref{sec:Results}.

\begin{figure*}[b]
    \includegraphics[width=\linewidth]{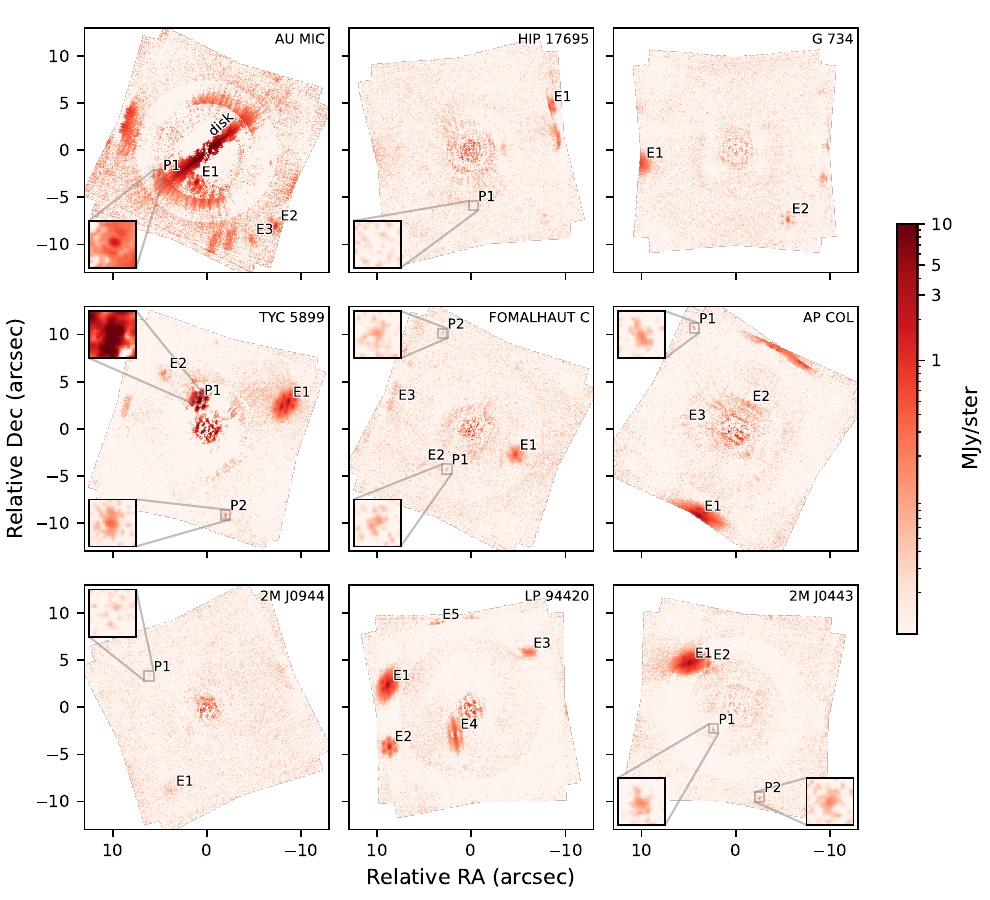}
    \caption{Flux maps of all targets in GTO 1184, in the F356W filter. All observations were reduced with RDI using the full library of reference PSFs. Point-like and extended sources are labeled in the images (with names beginning with ``P" and  ``E" respectively), and the point-like sources are magnified in the image insets. Extended background sources are abundant in the survey, and the edge-on disk of AU Mic is clearly detected.}
    \label{fig:flux_mosaic_f356w}
\end{figure*}

\pagebreak
\section{Individual Sensitivity Curves}
\label{a:contrast}

For brevity, we displayed the survey median contrast sensitivity curves in Section \ref{ssec:Survey_Sensitivity}, and here we show the results for each target individually. We achieve the background limited sensitivity at a separation of approximately 2.5" for each target, with the exceptions of AU Mic (due to its bright disk) and TYC 5899 (due to a bright off-axis source at $\sim$3"). The background-limited contrast is shallowest for the latest-type M dwarfs, as their intrinsic luminosities are the lowest while the background level is roughly constant across the targets. 

\begin{figure*}[h]
    \includegraphics[width=\linewidth]{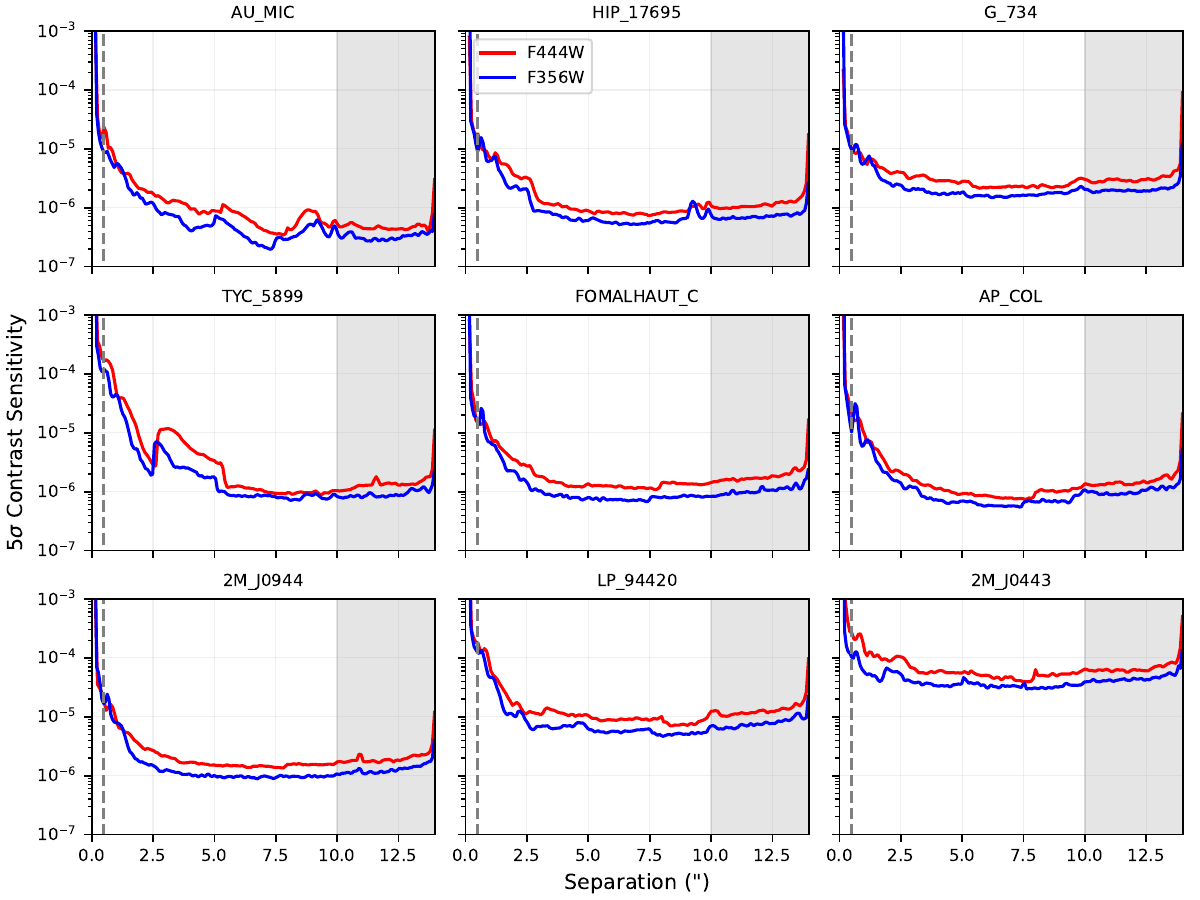}
    \caption{$5\sigma$ sensitivity curves for each target star in the F444W filter (red lines) and F356W filter (blue lines). The sensitivity is given in units of flux contrast in the lower panel. The coronagraph inner working angle (IWA) is shown in the gray dashed line, and the separations where only partial coverage exists due square framed observations at multiple roll angles is shown by the gray shaded region. We demonstrate extremely deep limits in absolute flux at close angular separations.}
    \label{fig:flux_contrast_mosaic}
\end{figure*}

\newpage

Here we show the 5$\sigma$ sensitivity limits in units of apparent magnitude for each target. We achieve a typical background-limited flux sensitivity of $\sim$ 21.5 in F444W and $\sim$ 22.5 in F356W. However the sensitivity is shallower for AU Mic due the use of the SHALLOW2 detector readout pattern for that target, as opposed to the MEDIUM8 pattern which was used for the rest of the targets.

\begin{figure*}[h]
    \includegraphics[width=\linewidth]{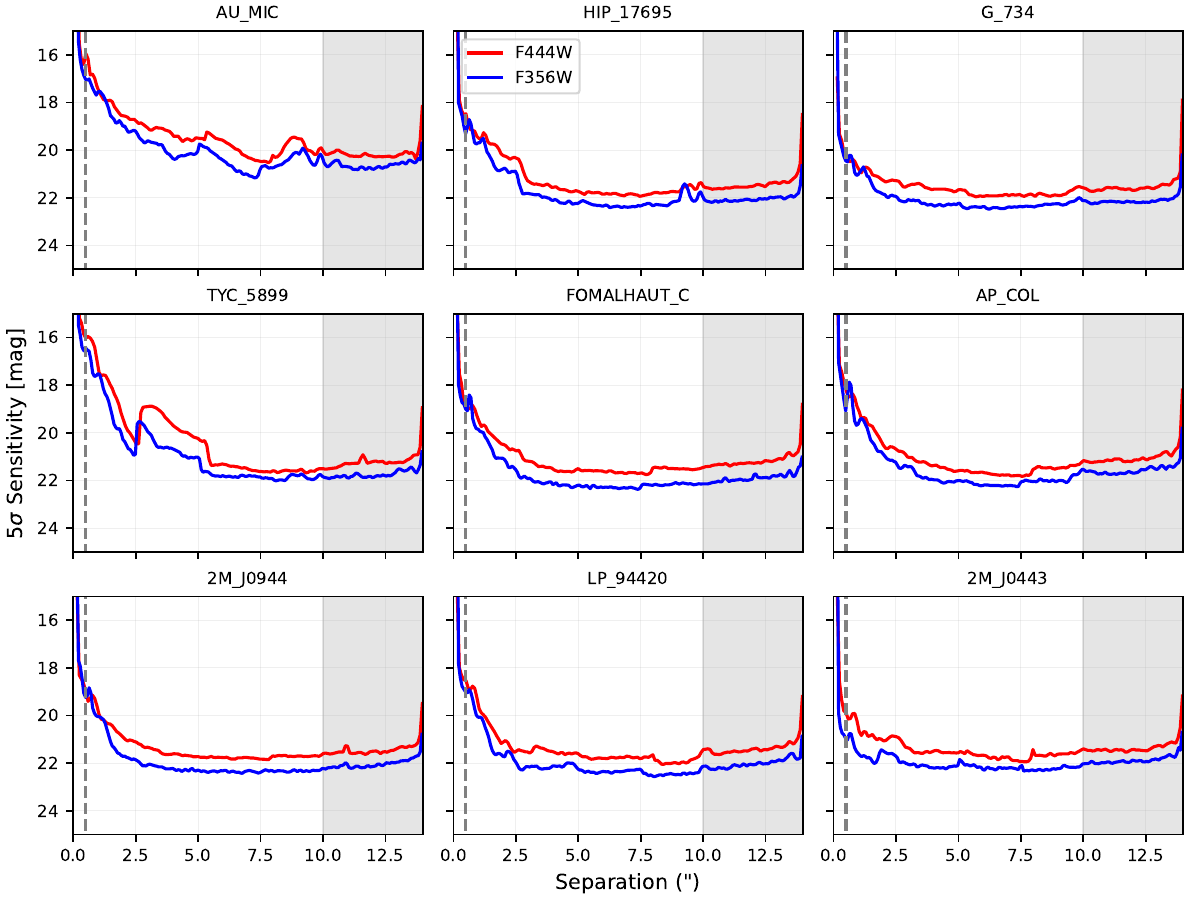}
    \caption{$5\sigma$ sensitivity curves for each target star in the F444W filter (red lines) and F356W filter (blue lines). The sensitivity is given in units of apparent magnitude in the lower panel. The coronagraph inner working angle (IWA) is shown in the gray dashed line, and the separations where only partial coverage exists due square framed observations at multiple roll angles is shown by the gray shaded region. We demonstrate extremely deep limits in absolute flux at close angular separations.}
    \label{fig:flux_sensitivity_mosaic}
\end{figure*}

\newpage
\section{Candidate Photometry}
\label{sec:FullPhotometry}

In Section \ref{ssec:Source_Catalog}, we show the best-fit parameters for each source in each filter using either the \code{spaceKLIP} or semi-independent least-squares PSF fitting methods, as occasionally the \code{spaceKLIP} method failed to converge or produceded erroneous results. Here we show the results for both methods for reference.

\begin{sidewaystable}[hb]
    
    \centering

    
    \begin{tabular}{lcr|rrrr|rrrr|l}
    \toprule
    \textbf{F444W} &  &  \multicolumn{1}{c|}{ } & \multicolumn{4}{c|}{spaceKLIP} & \multicolumn{4}{c|}{LSQ} \\
    Target & Source & SNRE & RA & Dec & Flux [mag] & SNR & RA & Dec & Flux [mag] & SNR & Method\\
    \cline{1-12}
    AU Mic & P1 & 7.21 & $5.057^{+0.014}_{-0.014}$ & $-2.51^{+0.011}_{-0.011}$ & $18.947^{+0.144}_{-0.144}$ & 7.68 & $5.052^{+0.009}_{-0.009}$ & $-2.511^{+0.008}_{-0.008}$ & $18.765^{+0.049}_{-0.049}$ & 7.59 & sklip \\
    \cline{1-12}
    HIP 17695 & P1 & 4.84 & - & - & - & - & $-0.281^{+0.007}_{-0.007}$ & $-5.837^{+0.009}_{-0.009}$ & $22.481^{+0.282}_{-0.282}$ & 1.64 & lsq \\
    \cline{1-12}
    TYC 5899 & P1 & 68.69 & $0.745^{+0.007}_{-0.007}$ & $3.031^{+0.007}_{-0.007}$ & $14.634^{+0.004}_{-0.004}$ & 75.11 & $0.748^{+0.007}_{-0.007}$ & $3.034^{+0.007}_{-0.007}$ & $14.644^{+0.007}_{-0.007}$ & 50.95 & sklip \\
     & P2 & 17.12 &  - & - & - & -  & $-1.979^{+0.008}_{-0.008}$ & $-9.107^{+0.007}_{-0.007}$ & $19.54^{+0.032}_{-0.032}$ & 12.97 & lsq \\
     \cline{1-12}
    Fomalhaut C & P1 & 9.03 & $2.521^{+0.058}_{-0.058}$ & $-4.219^{+0.112}_{-0.112}$ & $20.98^{+0.174}_{-0.174}$ & 6.83 & $2.489^{+0.032}_{-0.032}$ & $-4.197^{+0.009}_{-0.009}$ & $21.082^{+0.084}_{-0.084}$ & 4.98 & sklip \\
     & P2 & 5.76 & $2.992^{+0.042}_{-0.042}$ & $10.216^{+0.014}_{-0.014}$ & $21.153^{+0.108}_{-0.108}$ & 5.76 & $2.951^{+0.042}_{-0.042}$ & $10.203^{+0.014}_{-0.014}$ & $21.114^{+0.086}_{-0.086}$ & 4.67 & sklip \\
      \cline{1-12}
    AP Col & P1 & 5.81 & $4.385^{+0.037}_{-0.037}$ & $10.692^{+0.013}_{-0.013}$ & $21.042^{+0.125}_{-0.125}$ & 4.04 & $4.349^{+0.037}_{-0.037}$ & $10.7^{+0.068}_{-0.068}$ & $20.673^{+0.153}_{-0.153}$ & 4.85 & sklip \\
    \cline{1-12}
    2M J0944 & P1 & 4.02 & $6.098^{+0.018}_{-0.018}$ & $3.423^{+0.022}_{-0.022}$ & $21.842^{+0.093}_{-0.093}$ & 3.91 & $6.086^{+0.016}_{-0.016}$ & $3.402^{+0.022}_{-0.022}$ & $22.044^{+0.191}_{-0.191}$ & 2.47 & sklip \\
    \cline{1-12}
    2M J0443 & P1 & 9.28 & $2.314^{+0.016}_{-0.016}$ & $-2.264^{+0.011}_{-0.011}$ & $21.282^{+0.196}_{-0.196}$ & 4.08 & $2.311^{+0.011}_{-0.011}$ & $-2.259^{+0.011}_{-0.011}$ & $20.678^{+0.085}_{-0.085}$ & 5.94 & lsq \\
     & P2 & 8.44 & $-2.537^{+0.056}_{-0.056}$ & $-9.475^{+0.009}_{-0.009}$ & $20.865^{+0.095}_{-0.095}$ & 5.96 & $-2.593^{+0.056}_{-0.056}$ & $-9.481^{+0.009}_{-0.009}$ & $20.882^{+0.066}_{-0.066}$ & 6.66 & sklip \\

    \cline{1-12}
    \toprule
    \end{tabular}

    \begin{tabular}{lcr|rrrr|rrrr|l}
    \toprule
    \textbf{F356W} &  & \multicolumn{1}{c|}{ } & \multicolumn{4}{c|}{spaceKLIP} & \multicolumn{4}{c|}{LSQ} \\
    Target & Source & SNRE & RA & Dec & Flux [mag] & SNR & RA & Dec & Flux [mag] & SNR & Method\\
    \cline{1-12}
    AU Mic & P1 & 9.26 & $5.065^{+0.013}_{-0.013}$ & $-2.503^{+0.008}_{-0.008}$ & $19.203^{+0.118}_{-0.118}$ & 7.89 & $5.054^{+0.013}_{-0.013}$ & $-2.503^{+0.008}_{-0.008}$ & $19.006^{+0.051}_{-0.051}$ & 8.96 & sklip \\
    \cline{1-12}
    HIP 17695 & P1 & -0.39 &  - & - & - & -  &  - & - & - & -  & - \\
    \cline{1-12}
    TYC 5899 & P1 & 68.46 & $0.742^{+0.007}_{-0.007}$ & $3.033^{+0.007}_{-0.007}$ & $14.804^{+0.006}_{-0.006}$ & 68.02 & $0.745^{+0.007}_{-0.007}$ & $3.036^{+0.007}_{-0.007}$ & $14.735^{+0.01}_{-0.01}$ & 48.57 & sklip \\
     & P2 & 12.17 &  - & - & - & -  & $-1.982^{+0.008}_{-0.008}$ & $-9.102^{+0.008}_{-0.008}$ & $20.003^{+0.041}_{-0.041}$ & 11.38 & lsq \\
    \cline{1-12}
    Fomalhaut C & P1 & 5.41 & $2.483^{+0.018}_{-0.018}$ & $-4.184^{+0.013}_{-0.013}$ & $21.688^{+0.111}_{-0.111}$ & 5.84 & $2.506^{+0.013}_{-0.013}$ & $-4.182^{+0.012}_{-0.012}$ & $21.789^{+0.103}_{-0.103}$ & 4.59 & sklip \\
     & P2 & 4.59 & $2.996^{+0.031}_{-0.031}$ & $10.235^{+0.029}_{-0.029}$ & $21.997^{+0.157}_{-0.157}$ & 4.1 & $2.965^{+0.031}_{-0.031}$ & $10.206^{+0.029}_{-0.029}$ & $21.798^{+0.114}_{-0.114}$ & 4.38 & sklip \\
    \cline{1-12}
    AP Col & P1 & 5.39 & $4.345^{+0.019}_{-0.019}$ & $10.714^{+0.025}_{-0.025}$ & $21.828^{+0.215}_{-0.215}$ & 3.72 & $4.346^{+0.01}_{-0.01}$ & $10.74^{+0.028}_{-0.028}$ & $21.126^{+0.108}_{-0.108}$ & 4.60 & sklip \\
    \cline{1-12}
    2M J0944 & P1 & 2.98 &  - & - & - & -  &  - & - & - & -  & - \\
    \cline{1-12}
    2M J0443 & P1 & 6.45 &  - & - & - & -  & $2.299^{+0.092}_{-0.092}$ & $-2.262^{+0.047}_{-0.047}$ & $21.334^{+0.142}_{-0.142}$ & 5.84 & lsq \\
     & P2 & 7.99 & $-2.548^{+0.041}_{-0.041}$ & $-9.477^{+0.014}_{-0.014}$ & $21.444^{+0.083}_{-0.083}$ & 7.09 & $-2.589^{+0.041}_{-0.041}$ & $-9.49^{+0.014}_{-0.014}$ & $21.439^{+0.082}_{-0.082}$ & 5.99 & sklip \\

    \cline{1-12}
    \toprule
    \end{tabular}

    \caption{Photometric and astrometric results for each off-axis source, using both the spaceKLIP and LSQ PSF fitting methods as described in Section \ref{ssec:Image_Analysis}. The results for the F444W and F356W filters are shown in the upper and lower sub-tables respectively. We show the peak SNRE and preferred fitting method, as well as the best-fit relative right ascension, relative declination, flux, and PSF fit SNR using each method. In cases where a PSF fit failed, or was not performed for F356W dropout sources, a dash is displayed.}
\end{sidewaystable}

\newpage
\section{Extended Source Catalog}
\label{sec:ExtendedSources}

Here we show the list of extended sources identified in the GTO 1184 survey, with their approximate location and peak SNRE in each filter. While most of the sources are clearly visibly extended in the data, some were identified as extended only after the observation of a halo-like or extended structure in the residuals when fitting a point-like PSF.

\begin{table*}[h]
\centering
\begin{tabular}{ll|rr|rr}
\toprule
\textbf{Target} & \textbf{Source} & \multicolumn{2}{c|}{\textbf{Separation ["]}} &  \multicolumn{2}{c}{\textbf{Peak SNRE}}  \\
 &  & RA & Dec & F356W & F444W \\
\hline
\multirow[t]{3}{*}{AU MIC} & E1 & 1.038 & -3.241 & 25.62 & 21.38 \\
 & E2 & -7.394 & -7.960 & 19.47 & 16.16 \\
 & E3 & -4.751 & -9.408 & 4.00 & 6.85 \\
\cline{1-6}
HIP 17695 & E1* & -8.338 & 4.814 & 21.74 & 15.22 \\
\cline{1-6}
\multirow[t]{2}{*}{G 734} & E1 & 10.100 & -1.227 & 37.42 & 42.88 \\
 & E2 & -5.632 & -7.268 & 36.43 & 50.65 \\
\cline{1-6}
\multirow[t]{2}{*}{TYC 5899} & E1 & -8.716 & 2.989 & 172.44 & 197.97 \\
 & E2 & 4.499 & 6.073 & 11.88 & 17.25 \\
\cline{1-6}
\multirow[t]{3}{*}{FOMALHAUT C} & E1 & -4.688 & -2.674 & 39.28 & 40.45 \\
 & E2 & 5.192 & -3.744 & 4.95 & 5.91 \\
 & E3 & 8.275 & 2.674 & 1.92 & 7.35 \\
\cline{1-6}
\multirow[t]{3}{*}{AP COL} & E1 & 3.870 & -9.219 & 66.09 & 69.82 \\
 & E2 & -1.227 & 2.612 & 8.13 & 3.70 \\
 & E3 & 5.506 & 0.535 & 5.04 & -0.02 \\
\cline{1-6}
2M J0944 & E1 & 3.807 & -8.841 & 5.80 & 6.75 \\
\cline{1-6}
\multirow[t]{5}{*}{LP 94420} & E1 & 8.904 & 2.486 & 164.04 & 168.94 \\
 & E2 & 8.716 & -4.059 & 90.56 & 65.23 \\
 & E3 & -6.136 & 5.947 & 17.77 & 28.42 \\
 & E4 & 1.668 & -2.737 & 26.83 & 20.63 \\
 & E5* & 3.618 & 9.030 & 9.92 & 13.36 \\
\cline{1-6}
\multirow[t]{2}{*}{2M J0443} & E1 & 4.877 & 4.877 & 195.68 & 189.98 \\
 & E2 & 2.863 & 4.688 & 28.97 & 36.78 \\
\cline{1-6}
\toprule
\end{tabular}

\caption{We detected 22 extended sources across the 9 targets in the GTO 1184 survey, as labeled in Figures \ref{fig:Mosaic444}, \ref{fig:Mosaic356}, \ref{fig:flux_mosaic_f444w} and \ref{fig:flux_mosaic_f356w}. As our study is focused on point sources likely to be substellar companions, here we report only the approximate relative position to the target star and peak SNRE in F356W and F444W. The starred sources (HIP 17695 E1 and LP94420 E5) are affected by data wrapping at the edge of the field of view, and thus the relative position is likely erroneous.}
\label{table:ext_sources}
\end{table*}

\newpage
\begin{acknowledgements} \label{acknowledgements}

E.B. would like to thank the ExoExplorers program, which is sponsored by the Exoplanets Program Analysis Group (ExoPAG) and NASA’s Exoplanet Exploration Program Office (ExEP). 

This study has made use of data from the NASA Exoplanet Archive and the Mikulski Archive for Space Telescopes. This work was supported by the Sellers Exoplanet Environments Collaboration with funding from the NASA Internal Science Funding Model (ISFM) and the Center for Research and Exploration in Space Science and Technology II (CRESST II) Task 667.021. This work has also benefited explicitly from collaborations enabled by the Other Worlds Laboratory (OWL) at the University of California, Santa Cruz, a program funded by the Heising-Simons Foundation. M.D.F is supported by an NSF Astronomy and Astrophysics Postdoctoral Fellowship under award AST-2303911. Part of this work was carried out at the Jet Propulsion Laboratory, California Institute of Technology, under a contract with the National Aeronautics and Space Administration (80NM0018D0004).

Any opinions, findings, and conclusions or recommendations expressed in this material are those of the author(s) and do not necessarily reflect the views of the National Aeronautics and Space Administration.

\facility{JWST}

\software{Astropy \citep{astropy},
          Matplotlib \citep{matplotlib},
          Numpy \citep{harris2020},
          PHOENIX \citep{allard_2012},
          webbpsf \citep{perrin2012, perrin2014},
          pyKLIP \citep{wang_pyklip_2015, pueyo_detection_2016},
          jwst \citep{bushouse_jwst_2023},
          spaceKLIP \citep[][for this study, we used the specific version of \code{spaceKLIP} stored at \url{https://github.com/ell-bogat/spaceKLIP/commit/ca030ece6dd7ab9d567f56b808b424bd1ce5d702}]{kammerer_performance_2022},
          scikit-image \citep{scikit-learn},
          scipy \citep{scipy},
          synphot \citep{synphot},
          emcee \citep{emcee}}

\end{acknowledgements}

\bibliography{main}{}
\bibliographystyle{aasjournal}

\end{document}